\documentclass[article]{aa}  

\usepackage{graphicx}
\usepackage{multirow}
\usepackage{lineno}
\usepackage{xcolor}
\usepackage{soul}
\usepackage{placeins}
\usepackage{txfonts}
\begin{document}

   \title{Latitudinal variations in Neptune's temperature profile observed with ALMA}

   \author{\'Oscar Carri\'on-Gonz\'alez\inst{1,2}\thanks{\email{oscar.carrion@obspm.fr, oscar.carrion.gonzalez@gmail.com}}
         \and Raphael Moreno\inst{1}
         \and Emmanuel Lellouch\inst{1}
         \and Thibault Cavali\'e\inst{3,1}
         \and Sandrine Guerlet\inst{4,1}
         \and Gwena\"el Milcareck\inst{5}
         \and No\'e Cl\'ement\inst{6}
         \and J\'er\'emy Leconte\inst{3}
         }

   \institute{
        LIRA, Observatoire de Paris, Universit\'e PSL, CNRS, Sorbonne Universit\'e, Universit\'e Paris Cit\'e, 5 place Jules Janssen, 92195 Meudon, France
         \and
         Max-Planck-Institut f\"ur Astronomie, K\"onigstuhl 17, 69117 Heidelberg, Germany
         \and
         Laboratoire d’Astrophysique de Bordeaux, Univ. Bordeaux, CNRS, B18N, All\'ee Geoffroy Saint-Hilaire, 33615 Pessac, France
         \and
         Laboratoire de M\'et\'eorologie Dynamique/Institut Pierre-Simon Laplace (LMD/IPSL), Sorbonne Universit\'e, CNRS, \'Ecole Polytechnique, Institut Polytechnique de Paris, \'Ecole Normale Sup\'erieure (ENS), PSL Research University, Paris, France
         \and
         Instituto de Astrof\'isica de Andaluc\'ia (IAA), CSIC, Granada, Spain
         \and
         Institut Pierre-Simon Laplace, Sorbonne Universit\'e/CNRS, Paris, France}

 \abstract
{Despite the low solar irradiation it receives, Neptune shows a very active atmosphere with some of the most intense dynamics observed in Solar System atmospheres.
Characterizing the atmospheric temperature profiles of the planet is a key to understand these observed processes.
In this work, we derived the Neptune pressure-latitude thermal field, using 2016 ALMA measurements of the CO(3-2) spectral line at 345.796 GHz, with a spatial resolution of about 0.37" on Neptune's 2.24" disk. To analyse the data, we developed MCMC retrieval methods to derive both the temperature profiles and the CO abundance profile.
We find that our data probes the upper troposphere and the lower stratosphere of the planet, between 2\,bar and 0.1\,mbar. 

Although temperature and CO profile are strongly correlated, simultaneous retrievals of both parameters for disk-integrated observations reveal a factor of 2-3 larger CO abundance in the stratosphere than in the troposphere, reinforcing the hypothesis of CO delivery by a recent cometary impact. 
CO retrievals with fixed temperature profile do not fit the spatially-resolved observations, implying underlying temperature variations.
By performing temperature retrievals with spatially constant CO, we find distinct trends of the thermal profile between the southern polar region, mid-latitudes and the equator.
At 10--100\,mbar, this structure is consistent with that observed by Voyager 2 in 1989, i.e. colder mid-latitudes, with a warmer equator and south polar regions.
At 300-600\,mbar, however, we find a cold layer of about 45\,K at Southern polar regions (-80$^\circ$) which disappears towards mid-latitudes and the equator.

By probing pressure levels not easily accessible to other observing methods e.g. infrared sounding, submillimeter observations offer a new view of the complex thermal structure in the upper troposphere and lower stratosphere of the Icy Giants.}


   \maketitle
%

\section{Introduction} \label{sec:introduction}
The visit of the Voyager 2 spacecraft to Neptune in 1989 unveiled a very active atmosphere, with some of the most intense zonal winds in the Solar System \citep[e.g.][]{hammeletal1989, smithetal1989, sromovskyetal1993}.
Voyager also observed a banded structure, as well as bright and dark spots characteristic of different types of storms that modify the global reflectivity of the planet \citep[e.g.][]{hueso-sanchezlavega2019, huesoetal2020}.
This cloud activity has since then been monitored, revealing variability at the seasonal \citep{sromovskyetal2003, lockwood-jerzykiewicz2006, hammel-lockwood2007, karkoschka2011, huesoetal2017} and subseasonal \citep{sromovskyetal2001, irwinetal2016, wongetal2018, wongetal2022} time scales.
Whether the jets triggering these atmospheric dynamics are driven by shallow atmospheric processes \citep[e.g.][]{lian-showman2010}, or extend deeply into the planetary interiors and are powered by deep convection \citep[e.g.][]{suomietal1991, kaspietal2013, clementetal2024} remains unknown \citep[see e.g.][]{mouisisetal2018}.
In this context, characterising the thermal profile of Neptune's middle upper troposphere/lower stratosphere is a valuable path to understanding the observed dynamics of the planet.

Voyager 2 studied the thermal structure of Neptune 16 years before the Southern summer solstice of 2005, and revealed a warm pole and equator with cooler mid-latitudes at pressures (P) of 30mbar--1bar \citep{conrathetal1989, conrathetal1998}. 
From radio occultation measurements, \citet{lindal1992} found the temperature at the tropopause to be 52$\pm$2\,K, increasing then with altitude up to 130$\pm$12\,K at 0.3\,mbar.
\citet{bezardetal1991} used Voyager infrared observations of ethane and acetylene to probe the stratospheric (0.03-3\,mbar) thermal profile, and found the same trend observed at deeper pressure levels, i.e. cold mid-latitudes, with warmer equator and south pole.

Subsequent disk-averaged observations by the Infrared Space Observatory (ISO) in 1996/97 confirmed this temperature minimum of 53\,K at 150\,mbar \citep{burgdorfetal2003}.
From 1997 ISO data, \citet{bezardetal1999} found the upper stratosphere (20-0.3\,$\mu$bar) to be quasi-isothermal (165--180\,K), with a steep increase in temperature (T) above the 0.3\,$\mu$bar level reaching 250\,K at 0.1\,$\mu$bar and 550\,K in the thermosphere.
Stellar occultations have also been used to probe the upper atmosphere of Neptune, confirming the increase in temperature from tropopause levels to the thermosphere \citep{roquesetal1994, uckertetal2014}.
\citet{roquesetal1994} found temperatures of 150-200\,K at 25\,$\mu$bar from occultations between 1983 and 1990.
\citet{uckertetal2014} analysed a stellar occultation event from 2008 and found no temporal evolution from pre-Voyager data at P$\sim$10$\mu$bar.

Ground-based mid-infrared imaging has continued the monitoring of Neptune's upper troposphere, confirming the inhomogeneous thermal structure observed with Voyager.
From VISIR-VLT observations in 2006 --soon after the southern summer solstice--, \citet{ortonetal2007} found an enhanced thermal emission towards the south pole (derived temperature of 62-66\,K at 100\,mbar), possibly due to the recent maximum of solar irradiation.
\citet{fletcheretal2014} analysed mid-IR observations during Neptune's summer solstice --from 2003 to 2007-- and found no variations larger than 0.7\,K in the disk-averaged tropospheric zonal temperatures between Voyager's flyby and solstice.
From spatially-resolved data, they found a general trend in the latitude-dependent temperatures of the troposphere consistent with that of Voyager, with a minimum temperature at P=200\,mbar of 51\,K at southern (45$^\circ$) and northern mid-latitudes, and the equator and south pole at 56\,K.
This implies a warmer South polar region than in Voyager data.
They also found a hint of cooling towards -85$^\circ$, which they suggest could be due to spurious projection effects.
\citet{romanetal2022} reached similar conclusions about the stability of upper tropospheric temperatures, in this case with data from 2003 to 2020, and found a temperature of 49--52\,K at southern mid-latitudes, with 4--8\,K warmer equator and south polar regions.
They also observed a cooling of 6$\pm$3\,K in the south polar region between the 2003--2006 and the 2018--2020 epochs, at pressure levels of 50--300\,mbar.
This hints at the southern polar region being the region with greatest temporal variability in the troposphere.

In the stratosphere, \citet{ortonetal2007} found a distinct hot spot located at latitudes of 65$^\circ$--70$^\circ$ at  P=0.1\,mbar during solstice, although they noted that strong degeneracies between the temperature and the abundance of chemical species (CH$_4$ and C$_2$H$_6$) could affect this result.
With updated CH$_4$ abundances from Herschel \citep{lellouchetal2010}, \citet{fletcheretal2014} refined the quasi-isothermal stratospheric temperature profile from the Voyager-derived 165--180\,K by \citet{bezardetal1999} down to 158--164\,K.
Under the assumption of latitudinally uniform CH$_4$ abundance, they concluded that the stratospheric temperatures are also latitudinally uniform except for the warmer south polar vortex.
They also found that the stratospheric temperature variability between 1989 and solstice, if present, was less than $\pm$5\,K at 1\,mbar, and less than $\pm$3\,K at 0.1\,mbar.
Only the south pole experienced a temperature increase of 7-8\,K at 10-100\,mbar during the 1989-2005 time interval, with an increase of 5-6\,K over the broader southern 70$^\circ$-90$^\circ$ region at pressure levels of 0.1-200\,mbar.
With their extended 2003--2020 dataset, \citet{romanetal2022} reported a decrease in global stratospheric brightness after solstice, reaching a minimum in 2010, with hints of brightness peaks in 2012 and 2020.
This evidences subseasonal variations of the stratospheric temperatures, as opposed to their reported stability of the troposphere.
\citet{romanetal2022} also reported a new increase in stratospheric (0.01--2\,mbar) temperature at the south pole from 152$\pm$2\,K in 2018 to 163$\pm$3\,K in 2020.

Except for the Voyager radio occultations and a few stellar occultations, most previous temperature measurements rely on thermal IR sounding.
Yet, submillimeter observations, as exemplified by Herschel results on CH$_4$ \citep{lellouchetal2010, lellouchetal2015}, are also able to constrain the atmospheric temperatures.
Following the discovery of CO (and HCN) on Neptune at the James Clerk Maxwell Telescope operating in the submillimeter \citep{martenetal1993}, several authors acquired disk-averaged spectra of Neptune in CO(2-1) or CO(3-2).
For thermal sounding, however, a complication is the fact that the CO vertical profile is not vertically uniform.
From disk-averaged radio measurements, \citet{lellouchetal2005} found a non-uniform abundance of CO in the atmosphere, with a stratospheric abundance (1\,ppm at P<20mbar) twice that of the troposphere (0.5\,ppm at P>20mbar).
This CO abundance is an order of magnitude higher than that in Uranus \citep{encrenazetal2004, cavalieetal2014}, and potentially indicative of a relatively recent cometary impact that delivered CO to Neptune's stratospheric layers. 
This vertically inhomogeneous abundance of CO in Neptune was confirmed by additional radio measurements \citep{hesmanetal2007, luszczcook-depater2013} and data from the Herschel and AKARI space telescopes \citep{lellouchetal2010, lellouchetal2015, teanbyetal2019, fletcheretal2010}, spanning a time interval before and after the solstice.
Note that in these previous works, a temperature profile was assumed upon carrying CO abundance derivations, and the CO-temperature degeneracy was not addressed.

A new perspective on all these issues is now allowed by ALMA observations, which, compared to all previous heterodyne observations, adds the spatial dimension, since Neptune's $\sim$2.2" disk can be easily resolved \citep{tollefsonetal2019, tollefsonetal2021, iinoetal2018, iinoetal2020, carriongonzalezetal2023}. 
Previous CARMA interferometric observations, such as \citet{luszczcooketal2013}, also achieved spatial resolution on Neptune, although they could not measure the whole line profile but just a number of spectral windows at the core and wing of the CO line.

In this work, we explore the degeneracies between a priori unknown temperature and CO profiles by performing atmospheric retrievals using the CO(3-2) line.
The data were acquired with ALMA in 2016 (Sect. \ref{sec:observations}), during a period of time (November 2012 to August 2018) in which no mid-IR observations of Neptune are available \citep{romanetal2022}.
Given the sensitivity of our data to upper tropospheric and lower stratospheric levels, this provides complementary information to mid-IR imaging, and helps filling the gaps in the long-term monitoring of the planet after the southern summer solstice.
We developed atmospheric retrieval methodologies (Sect. \ref{sec:methods}) that allow us to simultaneously derive a temperature and a CO profile -- and that optionally allow to fix one of the profiles and only explore the other.
We present the retrieval results in Sect. \ref{sec:results}.
We first analyse disk-averaged data on which we perform temperature-CO joint retrievals --permitting us to revisit the problem of the CO vertical distribution, and then spatially-resolved observations from which we study the latitudinal variations of temperature.
In Sect. \ref{sec:discussion}, we discuss the results and summarize the main conclusions of this work.

\section{Observations} \label{sec:observations}

\begin{figure}
  	\centering
	\includegraphics[width=9cm]{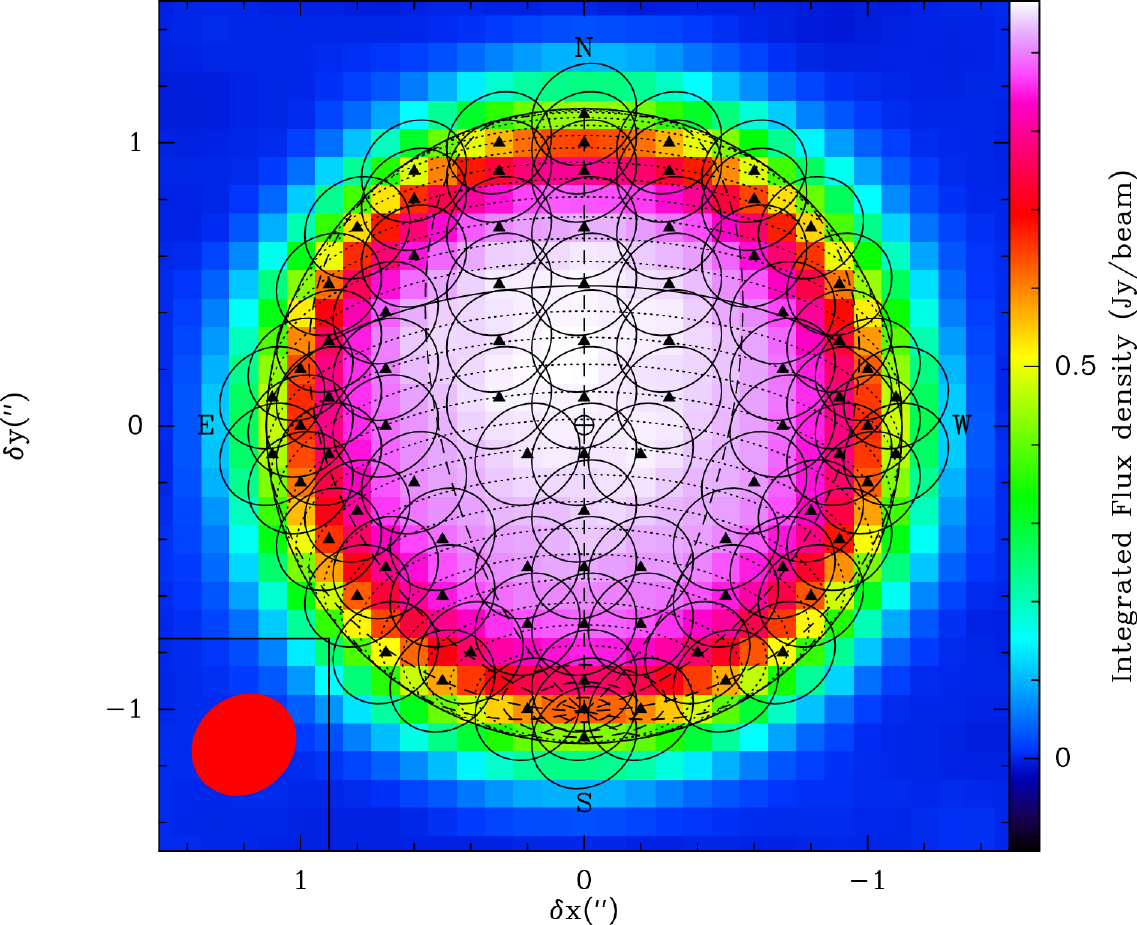} 
	\caption{Integrated flux for ALMA's CO(3-2) line measurement. Superposed as black circles on Neptune's map are the beam positions observed onto the planet and used for the spatially resolved retrievals. The filled red circle in the lower-left corner indicates the size and  orientation of the beam.}
\label{fig:observations}
\end{figure}

For this study we used the same ALMA dataset as in \citet{morenoetal2017} and \citet{carriongonzalezetal2023}.
These observations date from April 30, 2016 -- 11 years after the Southern summer solstice.
Data were acquired over an integration time of 20 minutes on-source, with the C36-2/3 hybrid configuration of the interferometer operating with 41 antennas.
The angular resolution achieved was about 0.37", enabling the distinction of several latitude bands as shown in Fig. \ref{fig:observations}.
The spectral resolution was 1\,MHz and the bandwidth, 4\,GHz.
Although the spectral setup included the HCN(4-3) line at 354.5054773 GHz, the CS(7-6) line at 342.883 GHz and the CO(3-2) line at 345.7959899 GHz, in this work we only focus on the analysis of the latter, leaving the analysis of the HCN and its 3D distribution for future work.

The calibration procedure for CO visibilities with the ALMA/CASA reduction software was the same as described in \citet{morenoetal2017} for CS.
The GILDAS package was used on the resulting calibrated visibilities, in the first place, to improve image quality by applying a self-calibration technique using Neptune’s continuum, and then to perform the imaging and deconvolution with the H\"ogbom algorithm \citep{hogbom1974}.
The absolute flux scale was derived from amplitude self-calibration using the \citet{lellouchetal2010} thermal profile and is accurate to within 5\%.
The resulting synthetic elliptical beam (robust weighting 0.5) was 0.39"$\times$0.34", with a polar angle of -47$^{\circ}$.
This resulted in a spatial resolution of about 20$^\circ$ at the equator of Neptune, which had an angular diameter of 2.24".
The final clean images were built with a sampling of 0.1" over $\{-1.4'',1.4''\}$.
The SNR at line peak at the limbs is about 150.

Figure \ref{fig:observations} shows the integrated flux of the CO(3-2) line recorded over Neptune's disk.
The ALMA beam is included for comparison, as well as the 83 spatially-resolved measurements used in our analysis.

\section{Methods} \label{sec:methods}
We built a retrieval methodology by combining a Markov chain Monte Carlo (MCMC) sampler with a radiative transfer code that produces synthetic spectra from input atmospheric models.

\subsection{Radiative transfer} \label{subsec:methods_RT}
\begin{figure}
  	\centering
    \includegraphics[trim = 1.25cm 2cm 4.75cm 2.5cm, clip, width=9cm]{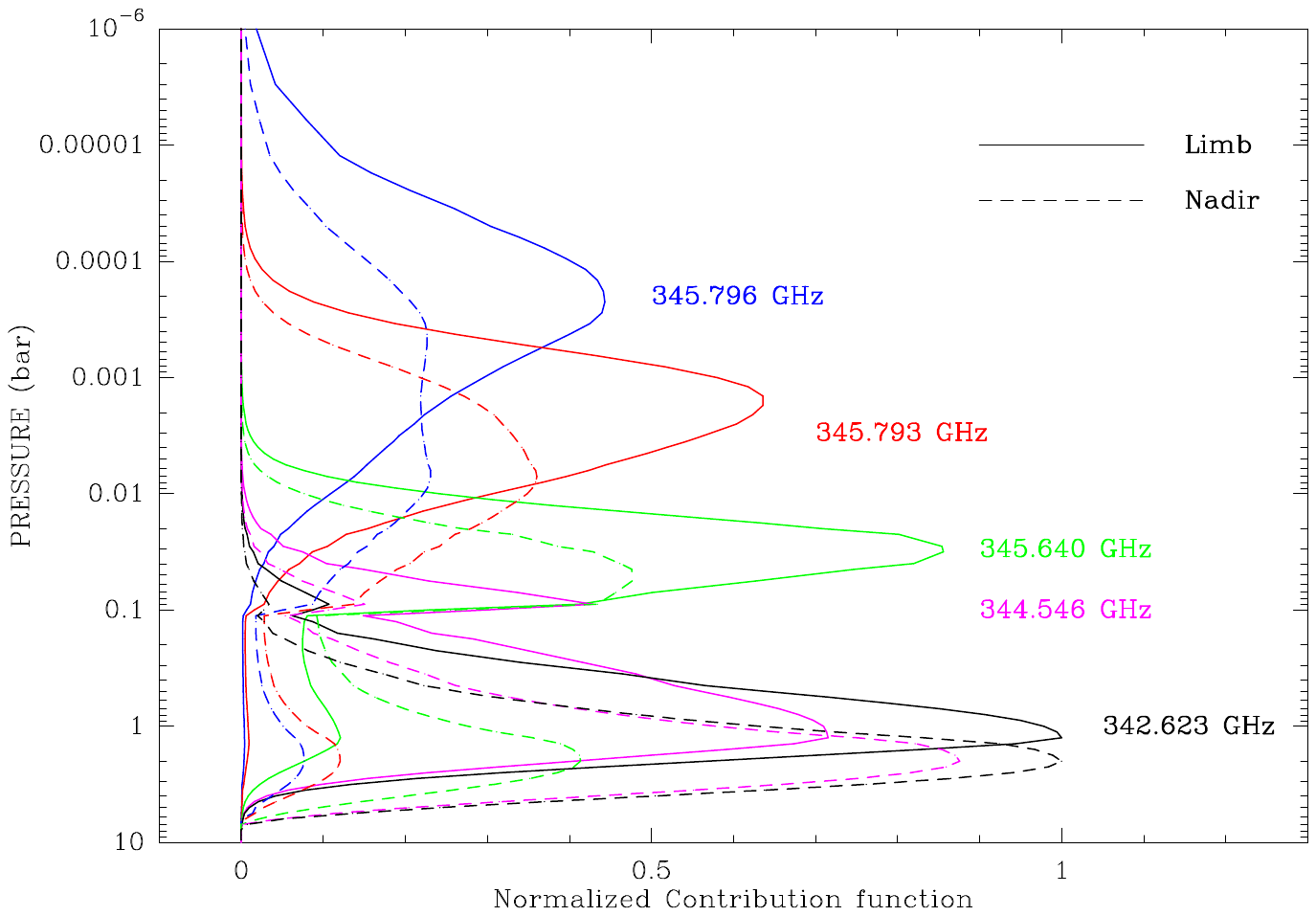} 
	\caption{Normalized contribution functions at a selection of frequencies for ALMA's CO(3-2) line measurement at Neptune's limb (solid lines) and nadir at the disk center (dashed lines).}
\label{fig:contribfunctions}
\end{figure}

We used the radiative transfer code from \citet{morenoetal2017}.
It takes as input both the local (beam-averaged) temperature profile and the vertical distribution of the relevant chemical species.
For this beam-averaging, the code includes pencil-beam calculations at all viewing geometries (including those beyond the limb), which are then averaged with proper weighting associated with the beam 2-D elliptical shape; this approach naturally accounts for beam dilution effects in limb observations.
The code considers a pressure range of about 100--$10^{-7}$\,bar, dividing the atmosphere into 128 slabs in logarithmic scale.
In our case, this enables simultaneous exploration of the temperature profile and CO abundance.
Separate retrievals of either temperature or CO can also be performed by keeping the other one fixed in the computations.
In either case, the output is a synthetic spectrum convolved by the observing beam and accounting for the viewing geometry and Neptune's solid-body rotation.
Winds, as measured by \citet{carriongonzalezetal2023}, are not explicitly included in the radiative transfer, but this is inconsequential for temperature/CO retrievals, given the 1 MHz resolution of the data.

Figure \ref{fig:contribfunctions} shows the beam-convolved contribution functions for a limb measurement, indicating the information contained at each channel of the spectrum, given by:
\begin{equation}
 \label{eq:contribfunct}
CF(\nu,z) = B(\nu,T)\cdot W(\nu,z)
\end{equation}
with the weighting function $W(\nu,z) = d(e^{(-\tau(\nu)})/dz$.
These were computed with our radiative transfer code (Sect. \ref{subsec:methods_RT}), and initially adopting the thermal profile and CO vertical distribution from \citet{lellouchetal2010}.
The continuum spectral frequencies probe pressure levels up to about 2\,bar, while the core of the CO(3-2) line can reach levels slightly above 0.1\,mbar.

\subsection{Atmospheric model} \label{subsec:methods_atmosmodel}
Building on previous results that point towards a dual source of CO, we described the atmospheric CO abundance as a step function consisting on three parameters: the tropospheric CO abundance (CO$_{tr}$), the stratospheric CO abundance (CO$_{str}$) and the pressure level at which the transition takes place ($P_0$).
This parameterization of Neptune's CO abundance is similar to that of e.g. \citet{lellouchetal2005, hesmanetal2007, fletcheretal2010, teanbyetal2019}.

For the temperature profile, we tested two different parameterizations: i) a polynomial function of T(P), and ii) a set of independent points (T$_i$ at some P$_i$ levels). 
The goal of this dual approach was to verify that the structure observed in the retrieved temperature profiles did not correspond to non-physical oscillations, e.g. due to the polynomial function.
To avoid blurring the complexity of the temperature profile by smoothing out potentially real structures, we did not apply any vertical smoothing to the temperature profiles generated during the retrieval process.
Our parameterization is thus the limiting factor for the vertical resolution; at the same time it prevents spurious small-scale structures in the retrieved profiles.

\subsubsection{Temperature polynomial parameterization} \label{subsubsec:methods_atmosmodel_polynomial}
In the first case, the temperature was parameterized as a $n^{th}$-order polynomial function of pressure between a given range of maximum and minimum pressure values -- generally, 5\,bar to 10$^{-5}$\,bar, encompassing with some margin the sensitivity range (Fig. \ref{fig:contribfunctions}).
Outside of this range, the temperature profile was forced to relax towards the reference profile from \citet{lellouchetal2010}.
This was needed to provide the whole pressure range used by the radiative-transfer code.
The parameters to be explored by the retrievals were then the $n+1$ polynomial coefficients.
We tested a variety of polynomial orders and pressure ranges for the temperature parameterization, adopting as our nominal case an $8^{th}$-order polynomial between 5 and 10$^{-5}$\,bar.

\subsubsection{Temperature independent-point parameterization} \label{subsubsec:methods_atmosmodel_points}
In the second case, we described the temperature profile by a set of $x$ independent points corresponding to the temperature at $x$ different pressure levels -- in this case: 800, 300, 100, 30, 10, 2, 0.2, and 0.02\,mbar.
Such a sampling with variations by a factor 2-3 of pressure between successive levels is appropriate given the width of the contribution functions (Fig. \ref{fig:contribfunctions}). 
Between each of these discrete points, the temperature was interpolated linearly in $log_{10}P$ to the pressure grid used in the radiative-transfer code.
Outside of this range, the temperature profile was also forced towards the reference profile from \citet{lellouchetal2010}.
This parameterization prevented any possible non-physical oscillations such as those that can could occur within the polynomial description, and was used as a way to double check the polynomial retrieval results.
The parameters explored in this second approach were directly the temperature values at each of the given pressure levels.

\subsection{Retrieval sampler} \label{subsec:methods_retrieval}
We used the \texttt{emcee} MCMC sampler \citep{foremanmackeyetal2013} to explore the space of parameters given by the atmospheric model. The uniform priors for the CO parameters were: [0.02, 2.0\,ppm] for both CO$_{tr}$ and CO$_{str}$, and [10,1000\,mbar] for $P_0$. For the $n^{th}$-order polynomial parameterization of temperature, the priors were [20,500] for the $0$-order polynomial coefficient and [-500,500] for the other $n$ coefficients. In the case of the independent-point temperature parameterization, the priors were set to $\pm 30$\% of the temperature reported at that pressure level in the reference profile of \citet{lellouchetal2010} -- e.g. $\pm$20\,K  at 800\,mbar, $\pm$16\,K at 100\,mbar, or $\pm$39\,K at the 2\,mbar level.
We compared this latter approach with a uniform set of wide temperature priors at each pressure level (e.g. [20,500\,K]), and found that exploring only the relevant $\pm$30\% temperature range optimised significantly the computation time while reaching equivalent results.

For each test atmospheric configuration sampled in the retrieval, the radiative-transfer code was run with the test CO and temperature profiles as inputs. The resulting spectrum (S$_{\text{test}}$) was compared to the ALMA measurement (S$_{\text{ALMA}}$) with the reduced $\chi^2$ figure of merit:

\begin{equation}
 \label{eq:chisq}
 \chi^2/N_\nu =
\sum_{i=1}^{N_\nu}\left(\frac{S_{test}-S_{ALMA}}{rms_{ALMA}}  \right)^2  / N_\nu
\end{equation}
with $N_\nu$ being the number of spectral data points (in this case, 3302 for the whole spectral range between 342.5 and 346 GHz), and $rms_{ALMA}$, the 1-$\sigma$ error of the measured data.
Throughout this work (e.g. Sect. \ref{subsec:results_diskav}) we will also refer to $\chi^2_{core}/N_{core}$, this being the equivalent to Eq. (\ref{eq:chisq}) but for a subset of $N_\nu$ corresponding to the $N_{core}$=72 spectral points at the emission line core (345.796$\pm$0.035\,GHz).

Eq. (\ref{eq:chisq}) computes $\chi^2/N$ only for a given observing geometry. 
That is, either for the disk-averaged spectrum, or for each of the measured spectra at the spatially-resolved beam positions observed onto the planet (Fig. \ref{fig:observations}).
In order to retrieve the average temperature profile at a given latitude --instead of at each beam position--, we ran multi-position retrievals.
In this case, for each test atmospheric configuration we ran the radiative transfer code at each of the beam positions located at that latitude.
The total reduced $\chi^2_{Lat}$ for that atmospheric configuration is then computed as the mean of the reduced $\chi^2_i$ at each beam position:
\begin{equation}
 \label{eq:chisq_multiple}
 \chi^2_{Lat} =
\frac{\sum_{i=1}^{n_{pos}}\chi^2_{i}/N_\nu}{n_{pos}}
\end{equation}
We note that the number of beam positions corresponding to a latitude ($n_{pos}$) changes significantly, from e.g. $n_{pos}$=11 at the equator to $n_{pos}$=1 for latitude -90$^\circ$.

For the polynomial retrievals, we used 500 chains ("walkers") to sample the parameter space while avoiding falling in local $\chi^2$ minima.
We observed that small variations in the coefficients of $8^{th}$-order polynomials produce large variations of the temperature profile, resulting in several of the walkers being lost during parts or all of the exploration process. 
This loss of walkers is due to non-physical temperature profiles, such as polynomials producing negative temperatures, unreasonably high temperatures that we discarded, or extreme oscillations of hundreds of K in adjacent atmospheric layers.
The independent-point parameterization retrievals, on the other hand, did not produce non-physical temperature profiles, and we used 20 walkers in this case. 
In all cases, we ran the retrievals for up to $10^6$ steps, and used a convergence criterion of 50 times the autocorrelation time \citep{foremanmackeyetal2013, goodman-weare2010}.
By comparing the retrieval results of both temperature parameterizations (Sect. \ref{sec:results}), we concluded that the sampling was successful in all cases.

The ensemble of good fits for each retrieval was defined as the 10\% best-fitting samples. We verified that the retrieved TP profiles and CO abundances remained practically the same if choosing either the 10\%, 5\% or 1\% best fits.
We interpret this as a validation that our convergence criterion ensured a thorough sampling of the region of the parameter space producing the best fits.
From our ensemble of good fits, we report both the best fit and the median value of each retrieved parameter, as well as its 1-$\sigma$ uncertainties.

\section{Results} \label{sec:results}
The effect of the temperature and CO profiles in producing the spectra are degenerate to some extent, which caused previous authors to fix the thermal profile upon analyzing disk-averaged spectra in terms of the CO vertical profile \citep{lellouchetal2005, hesmanetal2007, teanbyetal2019}.
Thus, we first explored this correlation by performing simultaneous retrievals of temperature and CO for the disk-average spectrum (Sect. \ref{subsec:results_diskav}).
We then analysed the spatially-resolved measurements (Sect \ref{subsec:results_spatialres}) and studied possible CO (Sect \ref{subsubsec:results_spatialres_COonly}) and temperature (Sect \ref{subsubsec:results_spatialres_Tonly}) spatial variations.

\subsection{Disk-averaged simultaneous retrievals of temperature and CO} \label{subsec:results_diskav}

\begin{figure}
  	\centering
    \includegraphics[width=9cm]{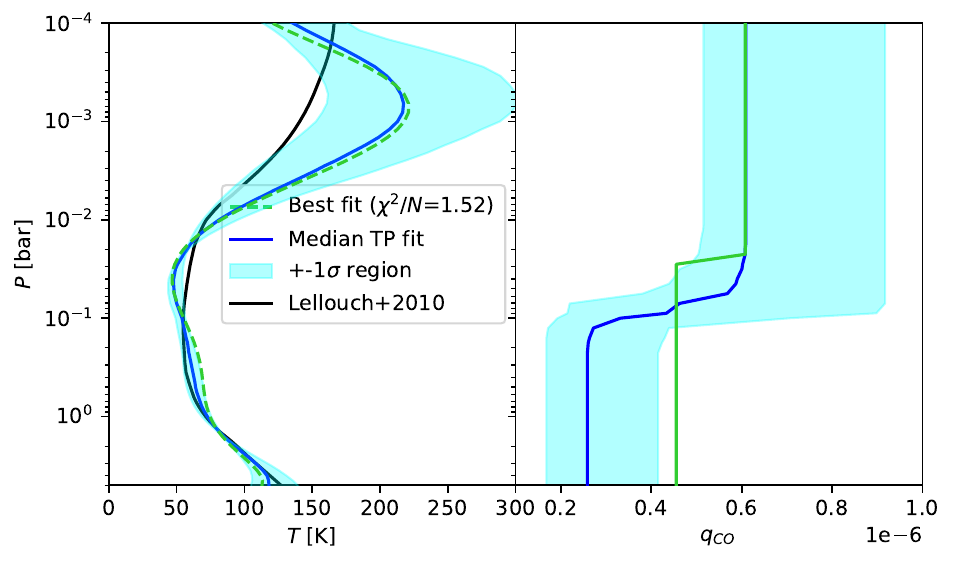}
    \\
  	\includegraphics[width=9cm]{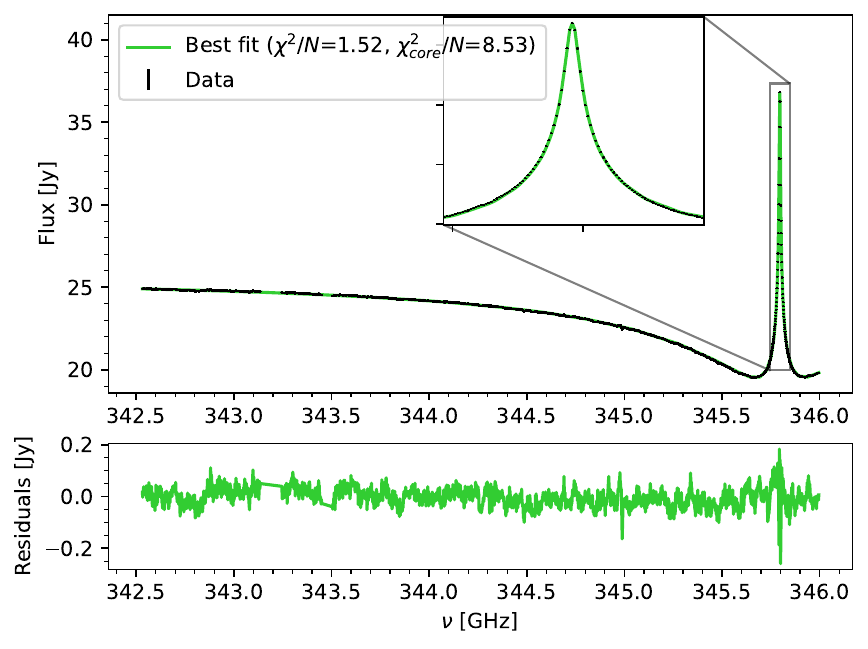} 
	\caption{Results of the disk-averaged retrieval exploring simultaneously temperature --parameterized as an 8th-order polynomial-- (top left), and CO abundance (top right). In blue are given the median of the ensemble of good fits, and green indicates the best-fit profile. Also included (bottom), the best-fitting synthetic spectrum from this retrieval, compared to the ALMA data.}
\label{fig:results_TP+CO_poly8}
\end{figure}

\begin{figure}
  	\centering
    \includegraphics[width=9cm]{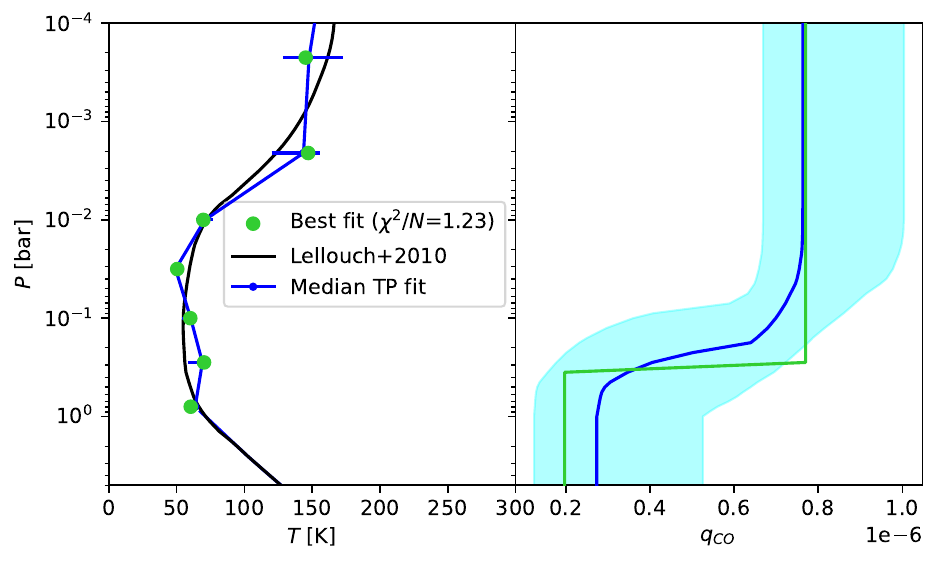}
    \\
  	\includegraphics[width=9cm]{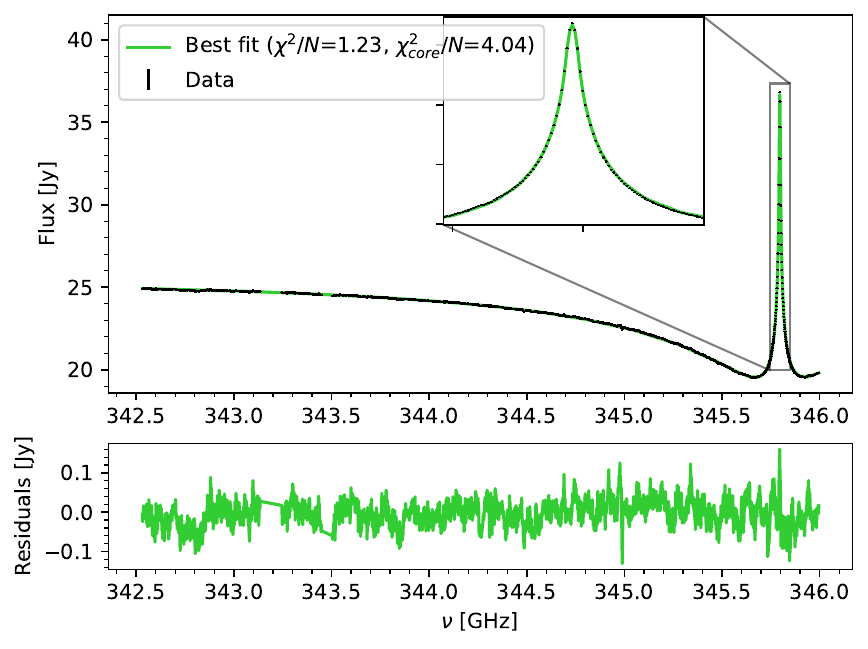} 
	\caption{As Fig. \ref{fig:results_TP+CO_poly8}, but here parameterizing the temperature as independent points.}
\label{fig:results_TP+CO_8point}
\end{figure}

We ran disk-average retrievals exploring simultaneously the CO abundance and the temperature profile as free parameters, both for the $8^{th}$-order polynomial parameterization of temperature (Fig. \ref{fig:results_TP+CO_poly8}) and for the independent-point parameterization (Fig. \ref{fig:results_TP+CO_8point}), and found consistent results in both cases.
The retrieved temperature profiles have a similar general structure, and both show a cold layer at about 30--40\,mbar: $48^{+7}_{-3}$K at 40\,mbar for the polynomial case, and $49^{+4}_{-3}$\,K at 32\,mbar for the independent-point parameterization.
This is significantly (about 2$\sigma$) colder than the reference profile from \citet{lellouchetal2010}, with 58\,K at 40\,mbar.

We also find that the independent-point parameterization performs better than the $8^{th}$-order polynomial at characterizing higher altitudes in the atmosphere.
In particular, at pressures lower than about 5\,mbar, the polynomial deviates towards warmer temperatures in a non-physical oscillation. 
We interpret this as an effect of the large changes in the temperature profile that are caused when the 9 polynomial coefficients undergo even small variations during the retrieval.
This effect --which is inherent to the polynomial parameterization and especially for high polynomial orders-- hinders the sampling of pressure levels to which our data is comparably less sensitive.

Indeed, pressure levels at <5\,mbar are probed at frequencies close to the core (345.796$\pm$0.035\,GHz) of the CO(3-2) emission line (Fig. \ref{fig:contribfunctions}).
This corresponds to only 72 spectral points out of the 3302 points of the complete spectrum.
The polynomial sampling thus focuses on fitting deeper levels, with more spectral points contributing to $\chi^2$.
The 72 points of the line's core have too low weight on $\chi^2$ to force the 9 polynomial coefficients to a different region of the parameter space, even more so in view of the large deviations of temperature --and hence of $\chi^2$-- produced by small changes in the coefficients.
The independent-point parameterization, on the other hand, is not affected by this issue, which allows the fitting of the line core to optimize in a continuous way the temperature determination at each level.
The values of $\chi^2_{core}$ in Figs. \ref{fig:results_TP+CO_poly8} and \ref{fig:results_TP+CO_8point} are in accordance with this interpretation.

A key result is that, for both parameterizations, the CO profiles show a difference of about a factor of two between the tropospheric and stratospheric abundances, with the transition at around 100\,mbar. 
From the ensemble of best fits, we retrieve for the polynomial temperature parameterization: CO$_{tr}=0.26^{+0.16}_{-0.09}$ppm, CO$_{str}=0.61^{+0.31}_{-0.09}$ppm, and $P_0=69^{+39}_{-39}$mbar.
Similarly, for the independent-point parameterization, we obtain: CO$_{tr}=0.27^{+0.25}_{-0.15}$ppm, CO$_{str}=0.76^{+0.24}_{-0.09}$ppm, and $P_0=202^{+328}_{-132}$mbar.
The distinct abundances of CO$_{tr}$ and CO$_{str}$ seem particularly robust given that the same priors were used for both parameters in the retrievals.
These results support the hypothesis of external CO delivery by a cometary impact as first presented in \citet{lellouchetal2005}.
We emphasize that our confirmation of a dual-source distribution of CO does not rely on any a priori assumption on the temperature profile, unlike in previous studies, and is therefore placed on even firmer grounds.
Nonetheless, the strong degeneracies between temperature and CO and the lower sensitivity in the stratosphere result in high uncertainties (30-50\%) in the retrieved CO parameters.
Therefore, the absolute values reported here for CO should be taken with caution.

As a way to further quantify the effects of the T--CO degeneracy on the temperature profiles, in Appendix \ref{sec:appendix_diskav_differentCO} we explored the effect of fixing different CO abundance profiles (i.e. varying CO$_{tr}$ and CO$_{str}$, keeping P$_0$ at 100\,mbar) and performing temperature-only retrievals on the disk-averaged data.
We find that the general structure of the temperature profile holds consistent in all cases (Fig. \ref{fig:appendix_TdifferentCO}), showing a cold layer near 40\,mbar.
Fig. \ref{fig:appendix_TdifferentCO} shows that a difference in the tropospheric CO of $\Delta(CO_{tr})$=20\% results in a change in the tropospheric temperature of $\Delta(T)\sim5\%$.
On the other hand, $\Delta(CO_{str})$=20\% produces a variation of $\Delta(T)\sim10\%$ in the stratospheric temperature.

\subsection{Spatially-resolved retrievals} \label{subsec:results_spatialres}
\subsubsection{CO spatial variations?} \label{subsubsec:results_spatialres_COonly}

We first attempted to fit the observed spectra at each of the latitudes --between -90$^\circ$ and +53$^\circ$-- by fixing the temperature profile to that of \citet{lellouchetal2010} and only exploring the CO parameters (Appendix \ref{sec:appendix_CO-only}).
We found that variations of up to a factor of three in the retrieved CO$_{tr}$ were needed to fit the observations at different latitudes, and up to a factor of two for CO$_{str}$ and $P_0$ (Table \ref{table:appendix_CO-only}).
Except in the case of a very recent (typically less than 20 years old, based on Jupiter Shoemaker-Levy 9 analogy) cometary impact that can temporarily induce a spatially variable CO distribution, a non-condensible and chemically long-lived species such as CO is expected to be uniformly distributed throughout the planet. 

Since CO is vertically not uniformly mixed due to its dual external/internal origin, however, a departure from this situation could occur if its horizontal distribution is affected by latitude-dependent upwelling/downwelling motions. An example is provided by the non latitudinally uniform distribution of PH$_3$ in Jupiter and Saturn, which results from the balance between photochemical depletion and vertical transport, either through meridional overturning or eddy mixing \citep{fletcheretal2009}. Here, the retrieved variability in the CO parameters, and particularly the P$_0$ value, which shows a minimum at low latitudes, could be interpreted as due to vigorous upward motion at low latitudes in the upper trosposphere, pushing P$_0$ to lower pressure levels. However, this scenario is at odds with recent views of the circulation in Neptune's troposphere, which call for stacked circulation cells with, in the equatorial region, a P>1\,bar upwelling cell meeting a P<1\,bar subsidence cell near the CH$_4$-ice cloud tops \citep[see e.g. Fig. 5 of][]{fletcheretal2020}.

Furthermore, these CO retrievals with horizontally uniform thermal profile do not allow a satisfactory fit of the measured data. 
Table \ref{table:appendix_CO-only} shows the reduced $\chi^2$ values for each of these retrievals.
In general, these values are significantly larger than $\chi^2/N\sim1$, which is the usual criterion for a good fit.
For comparison, Table \ref{table:appendix_CO-only} includes the reduced $\chi^2$ values for the temperature-only retrievals with a fixed CO profile (CO$_{tr}$=0.2\,ppm, CO$_{str}$=1.0\,ppm, and $P_0$=100\,mbar, see Sect. \ref{subsubsec:results_spatialres_Tonly}).
These temperature retrievals produce much better fits to the data.
Figure \ref{fig:appendix_onlyCO} shows, for an example latitude of -80$^\circ$, the difference between the measured data and the best-fit synthetic spectra for both CO-only and T-only retrievals, clearly illustrating the superiority of the latter case.

\citet{luszczcooketal2013} carried out a similar exercise for CARMA band measurements of Neptune in the CO(2-1) line.
They could not reproduce the spatially-resolved data with horizontally uniform CO and thermal profiles.
They found that a uniform CO profile with latitudinal temperature variations reproduced the observations with the best-fitting $\chi^2$ value.
An alternative model with uniform temperature and higher CO abundance in the mid-latitudes than everywhere else was also compatible with the data, although with worse values of $\chi^2$ \citep{luszczcooketal2013}.
In our case, the CO-only retrievals with fixed temperature (Table \ref{table:appendix_CO-only}) not only have overall worse $\chi^2$ than the T-only retrievals with CO fixed.
But, in addition, they suggest an enhanced CO abundance at the equator with respect to mid-latitudes --the opposite of the trend retrieved by \citet{luszczcooketal2013}.
We consider such short-scale temporal variations of the latitudinal CO distribution extremely unlikely.
Overall, this exercise, along with previous results from \citet{luszczcooketal2013}, evidenced underlying variations in the latitudinal thermal profiles, which we study below.
This is consistent with the evidence from previous mid-infrared analyses and the conclusions therefrom \citep[e.g.][]{fletcheretal2014, romanetal2022}.

\subsubsection{Temperature spatial variations} \label{subsubsec:results_spatialres_Tonly}

\begin{figure}
  \centering
  \includegraphics[width=8.5cm]{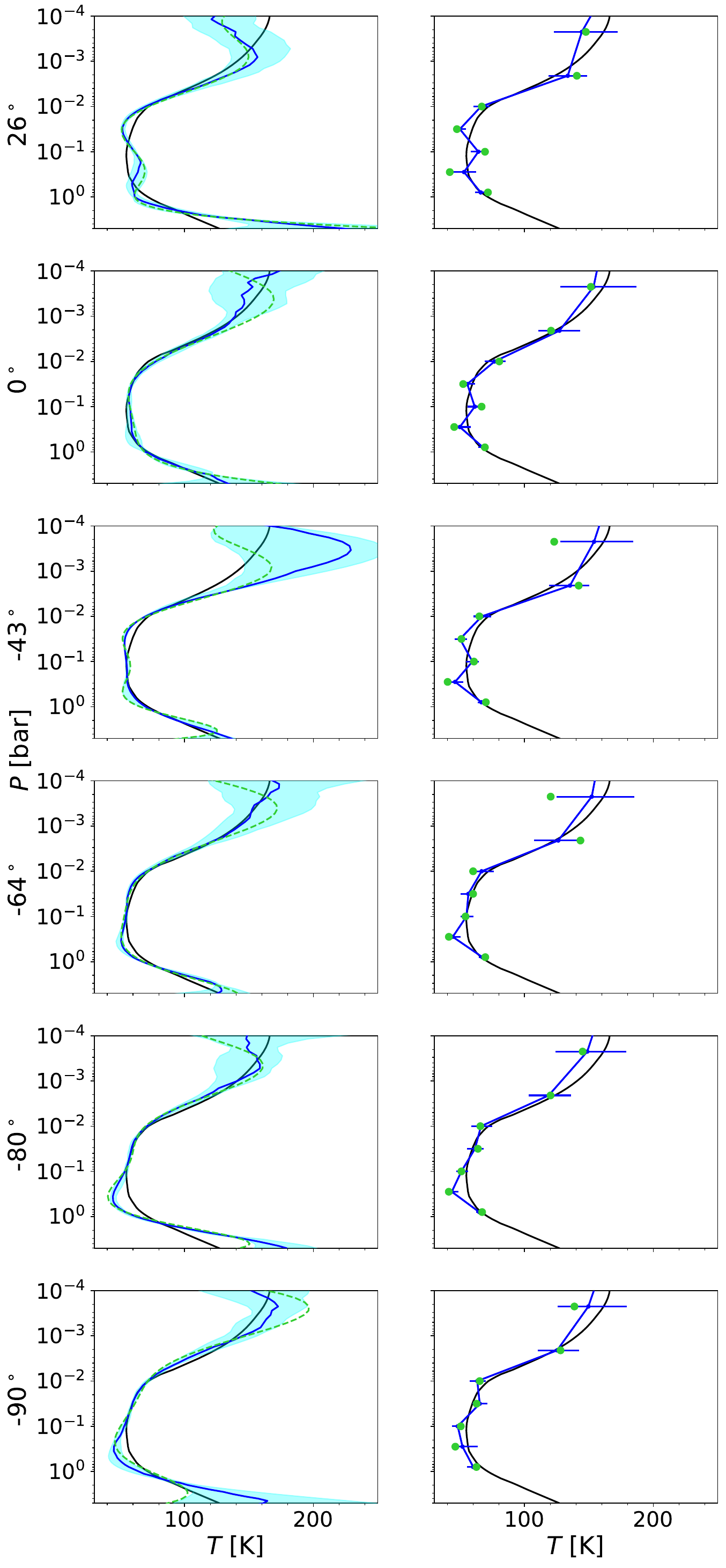}
    \caption{Results of the spatially-resolved temperature-only retrievals at different representative latitudes, for both the polynomial parameterization (left column) and the independent-point parameterization (right).
    As in Figs. \ref{fig:results_TP+CO_poly8} and \ref{fig:results_TP+CO_8point}, blue lines indicate the median of the ensemble of good fits, green lines (left) or dots (right) represent the best-fit profile, and black lines show the \citet{lellouchetal2010} profile for reference.}
\label{fig:temperatures_multiplelatitudes}
\end{figure}

\begin{figure}
  \centering
  \includegraphics[width=8.25cm]{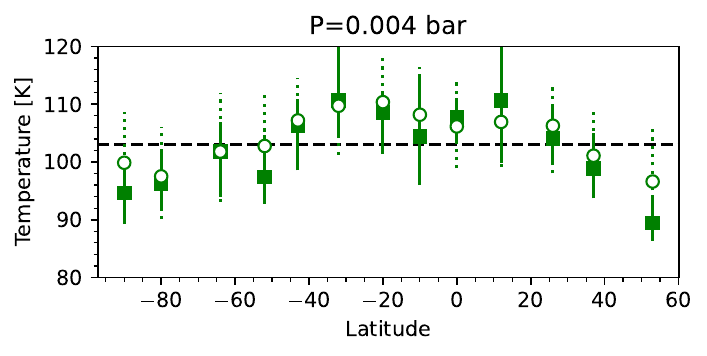}
  \vspace{-0.5cm}
  \\
  \includegraphics[width=8.25cm]{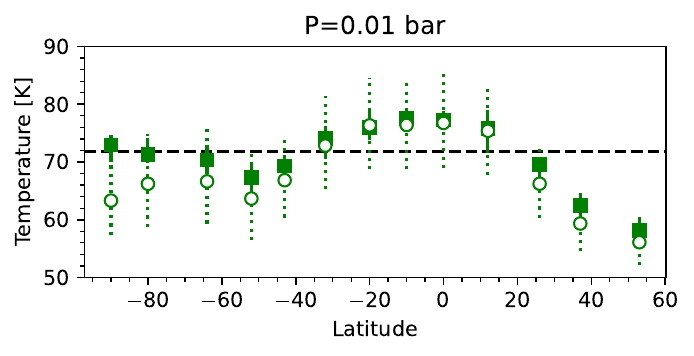}
  \vspace{-0.5cm}
  \\
  \includegraphics[width=8.25cm]{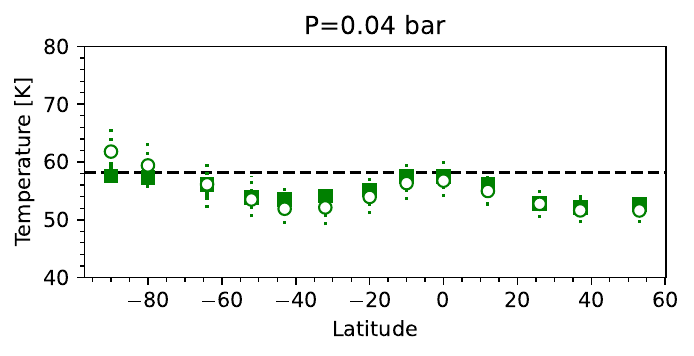}
  \vspace{-0.5cm}
  \\
  \includegraphics[width=8.25cm]{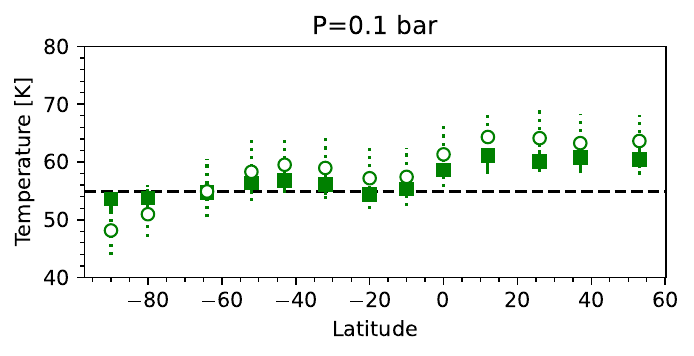}
  \vspace{-0.5cm}
  \\
  \includegraphics[width=8.25cm]{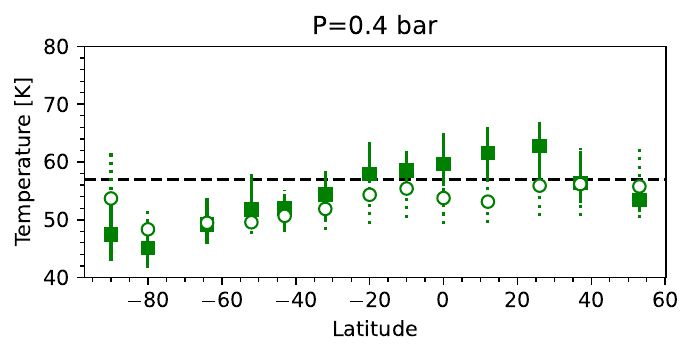}
  \vspace{-0.5cm}
  \\
  \includegraphics[width=8.25cm]{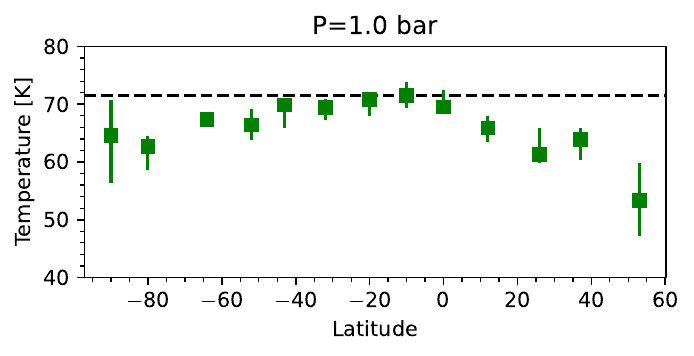}
  \vspace{-0.3cm}
  \caption{Latitudinal temperature variations at several pressure levels from the spatially-resolved retrievals, with a polynomial parameterization (square markers and solid error bars), and with the independent-point one (circles and dashed error bars, except at 1 bar). The median temperature retrieved and their uncertainties are shown, compared to that from \citet{lellouchetal2010} at that pressure (horizontal line). }
\label{fig:Tlatitude-cuts}
\end{figure}

\begin{figure}
  \centering
  \includegraphics[width=9cm]{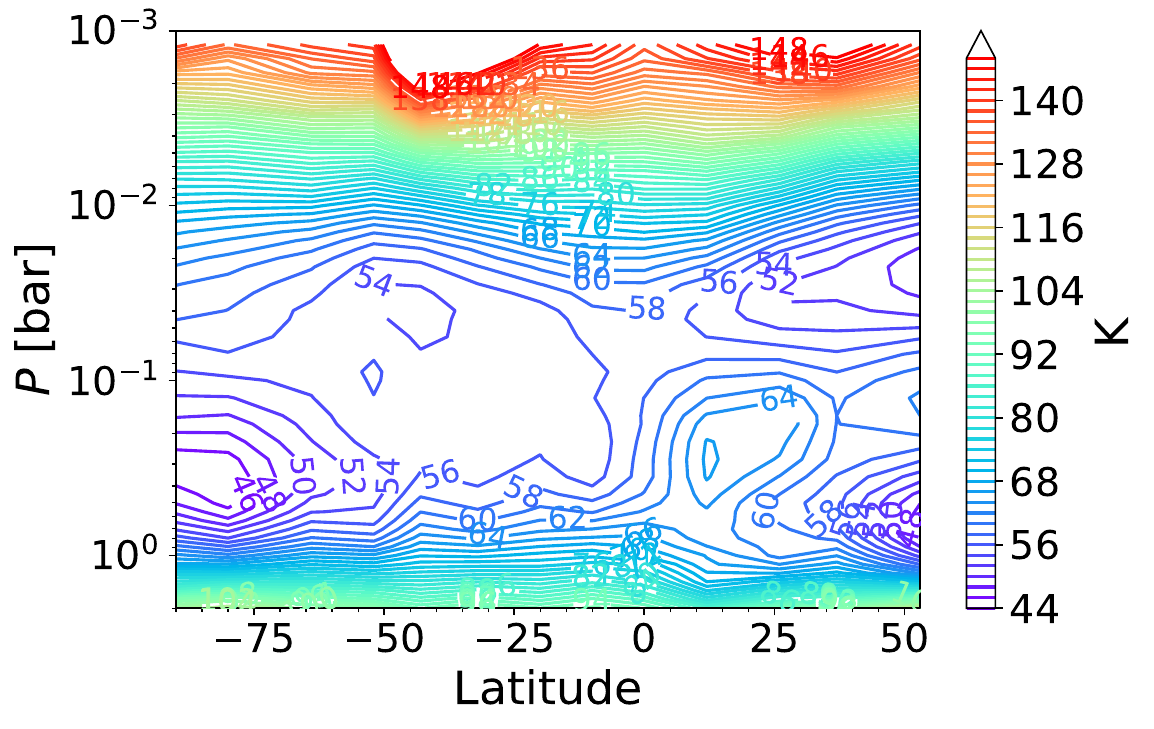}
  \\
  \includegraphics[width=9cm]{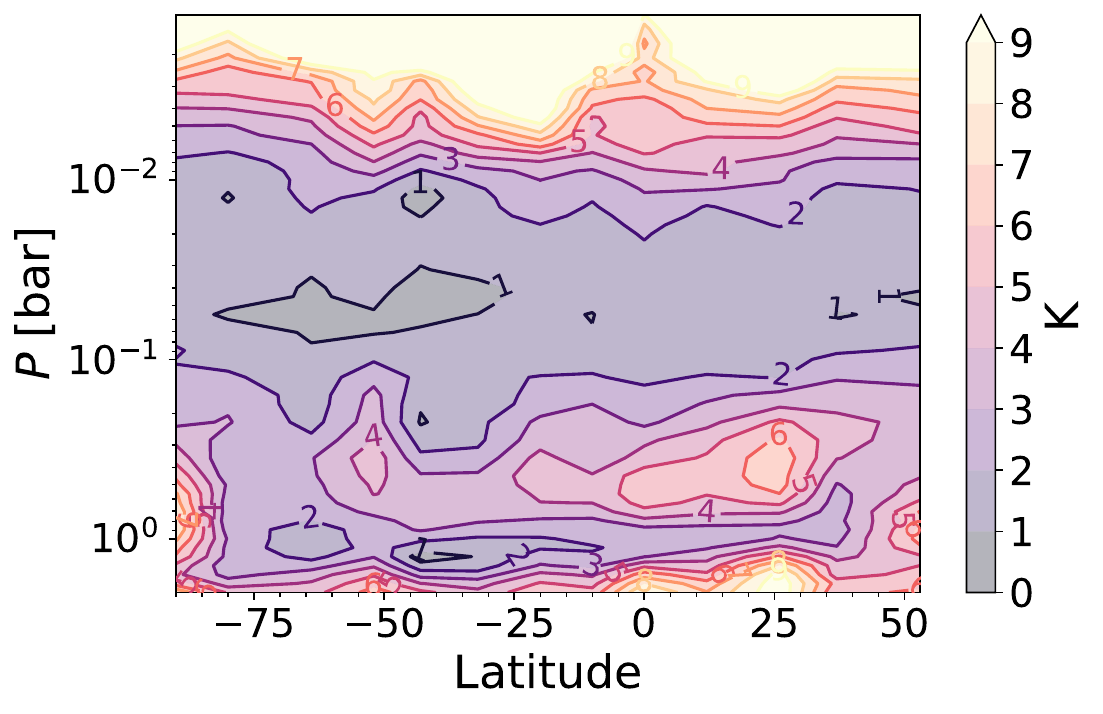}
  \\
  \includegraphics[width=9cm]{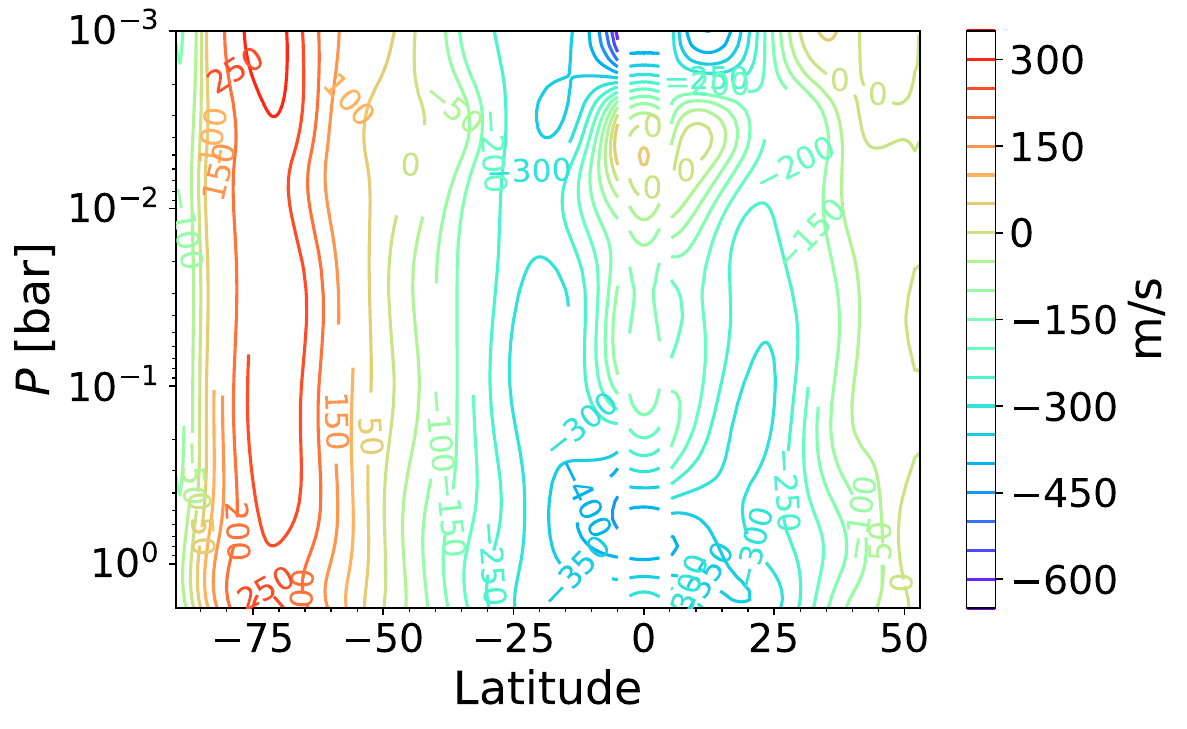}
    \caption{Temperature and wind maps built from the spatially-resolved retrievals, with a polynomial temperature parameterization. The top panel shows the median temperature profile retrieved at each pressure level and latitude. The middle panel shows the corresponding uncertainty of these results. The bottom panel corresponds to the thermal winds computed as in \citet{benmahietal2021}.}
\label{fig:Tmaps}
\end{figure}

To study the latitudinal temperature variations, we fixed the CO profile in all cases to CO$_{tr}$=0.2\,ppm, CO$_{str}$=1.0\,ppm, and $P_0$=100\,mbar.
These CO abundances are motivated by the results of the aforementioned CO-only retrievals (Table \ref{table:appendix_CO-only}), in particular those at low latitudes, which dominate the projected area of Neptune, and where the $\chi^2$ values are best.
The adopted abundances are also consistent within error bars to those retrieved from the T+CO disk-average retrievals from Sect. \ref{subsec:results_diskav}, and with previous works in the literature \citep{lellouchetal2005, teanbyetal2019}.

We note however that the partial degeneracies between temperature and CO prevented us from obtaining accurate absolute values for the CO parameters in Sect. \ref{subsec:results_diskav}, and in particular for the stratospheric levels which bear relatively less weight in the spectrum -- as discussed above.

Figure \ref{fig:temperatures_multiplelatitudes} shows the retrieved temperature profiles at a selection of representative latitudes, and Fig. \ref{fig:appendix_retrievals8pointVS8order} contains all 13 latitudes under study.
Latitudinal profiles of temperatures at a few selected pressure levels are shown in Fig. \ref{fig:Tlatitude-cuts} for both temperature parameterizations -- except at 1\,bar, given that temperature was not sampled at such pressure level in this parameterization.
The temperature-latitude field is presented in Fig. \ref{fig:Tmaps} (first panel), along with its error bar (second panel). The last panel of Fig.\ref{fig:Tmaps} shows thermal winds derived from this thermal field, using the methodology of \citet{benmahietal2021}, which allows thermal winds to be computed even at low latitudes, following the equations from \citet{marcusetal2019}. In these calculations, the zonal winds measured by Voyager at 1\,bar \citep{sromovskyetal1993, sromovskyetal2001b} were used as a boundary condition.
Table \ref{table:appendix_CO-only} (last column) shows the $\chi^2$ values of the best-fitting model at each latitude. 
As detailed hereafter, we find a clear structure in the latitude-dependent temperature profiles, with a continuous and consistent evolution from pole to pole.
These retrieval results are equally observed either using the polynomial or the independent-point temperature parameterization.

Although here we focus on the results for our adopted CO profile, we note that we repeated this exercise with other fixed CO profiles (as in Fig. \ref{fig:appendix_TdifferentCO}, but for the spatially-resolved retrievals) and found the same general structure of the thermal field.

At Southern polar regions, in particular at -80$^\circ$, we find a cold layer of 45$\pm$3\,K at about 400\,mbar. 
This cold layer is progressively subdued both at -90$^\circ$ and towards Southern mid-latitudes -- e.g. at -64$^\circ$.
Additional observations with higher spatial resolution are however needed to refine the characterization of this cold spot, as the beam in our observations (Fig. \ref{fig:observations}) is not small enough to isolate the South pole without the influence of nearby regions.
Towards mid-latitudes, we also observe a cold region at layers between 20 and 40\,mbar, consistent with the cold layer at 40\,mbar observed in the disk-average retrievals (Figs. \ref{fig:results_TP+CO_poly8} and \ref{fig:results_TP+CO_8point}).
This results in a slightly two-peaked temperature profile at Southern mid-latitudes ($-52^\circ$,  $-43^\circ$,  $-32^\circ$), with local temperature minima of about 53\,K at both 40\,mbar and 400\,mbar, and temperatures at 100\,mbar similar to those in \citet{lellouchetal2010}.
Although the contribution functions in Fig. \ref{fig:contribfunctions} show high sensitivity to both 40\,mbar and 400\,mbar levels, the radiances are less sensitive to temperature conditions near 100\,mbar, in relation with the drop-off of CO at higher pressures.
We thus take this subtle zig-zag shape --observed for both temperature parameterizations, and which has not been reported before-- with caution. 
At equatorial regions ($-20^\circ$,  $-10^\circ$,  $0^\circ$, see Fig. \ref{fig:appendix_retrievals8pointVS8order}), the retrieved temperature profile goes back to practically that from \citet{lellouchetal2010}, and shows no sign of the 20--40\,mbar temperature minimum found at other latitudes.
Finally, at Northern mid-latitudes we retrieve a two-peaked temperature profile even more marked than that in the South, although the uncertainties are also higher (see e.g. Figs. \ref{fig:temperatures_multiplelatitudes} and \ref{fig:appendix_retrievals8pointVS8order}).
We note that due to the observing geometries and associated projection effects, the characterization of Northern latitudes is inherently more uncertain.
This happens because each beam contains a wide range of latitudes, potentially with different temperature profiles, and this effect is more pronounced for observations of the Northern hemisphere.

In all cases, retrieval results at pressures lower than 1\,mbar become significantly more uncertain.
As discussed above, this is especially the case for polynomial retrievals due to non-physical oscillations of the function.
Independent-point temperature retrievals, on the other hand, are not affected by this issue and their main source of uncertainty at <1\,mbar pressure levels is the relatively small weight of the emission core in the overall $\chi^2$.
Similarly, the polynomial fits e.g. in Fig. \ref{fig:temperatures_multiplelatitudes} show a non-physical behaviour in the lower troposphere, towards 5\,bar. 
We carried out multiple tests with different maximum pressures explored in the retrievals, and found that this is due to the lack of sensitivity at those deeper layers.
Nonetheless, the south polar cold peak at 400\,mbar is observed in all cases. 
This, together with the detection of the cold peak with the independent-point temperature parameterization,  ensures that the cold peak is not an artifact of the polynomial retrievals.

\section{Discussion and conclusions}    \label{sec:discussion}
We have used ALMA to characterize the atmosphere of Neptune with spatially-resolved measurements of the CO(3-2) line.
For this, we developed retrieval methodologies, using two different parameterizations of the thermal profile, both capable of exploring the CO profile as well.
This 2016 dataset helps to fill a gap in ground-based mid-IR imaging (no data available from 2012 to 2018), and probes pressure levels that were not easily accessible to previous works.

Indeed, our measurements probe pressures between 0.1\,mbar and 2\,bar, with the contribution functions peaking both between 0.1 and 70\,mbar, and between 200\,mbar and 2\,bar (Fig. \ref{fig:contribfunctions}).
This pressure range is complementary to previous works observing in the mm (ALMA) and cm (VLA) continuum emission \citep{depateretal2014, tollefsonetal2019, tollefsonetal2021}, which were sensitive to deeper layers: 1 to 10\,bar, and 1 to 50\,bar, respectively.  
It is also highly complementary to mid-IR imaging data probing higher up in the atmosphere \citep{fletcheretal2014, romanetal2022}.
For instance, the seven filters (8--18.8\,$\mu$m) used by \citet{fletcheretal2014} for their retrievals had three sensitivity peaks at about 0.1 to 1\,mbar, 3 to 6\,mbar, and 40 to 200\,mbar \citep[][Figs. 13 and 14 therein]{fletcheretal2014}. 
Similarly, the contribution functions in \citet[][Figs. 20 and 21 therein]{romanetal2022} showed two peaks, at 0.1-2\,mbar and 50-200\,mbar respectively, but the sensitivity decreased significantly between about 2 and 50\,mbar, and also for P>300\,mbar, dropping to zero for P>500\,mbar due to the onset of the 10--20\,$\mu$m continuum.
One asset of the mm/submm range is that the weakness of the H$_2$-He collision-induced absorption allows probing down to the $\sim$2 bar level and that, given the nearly linear temperature dependence of the radiances, strong sensitivity is retained even to cold atmospheric layers.

Our high-SNR disk-integrated measurements allowed us to perform simultaneous retrievals of the global CO abundance and thermal profile.
Although we find strong correlations between the temperature and CO profiles in the retrieval, which leads to large error bars when the two parameters are set free, we obtain consistent thermal profiles with both temperature parameterizations.
In both cases, we find a cold layer of about $49^{+5}_{-3}$\,K  at a pressure level of 40\,mbar (Figs. \ref{fig:results_TP+CO_poly8}, \ref{fig:results_TP+CO_8point}).
The uncertainties in the absolute value of the temperature at 40\,mbar are large due to the T-CO degeneracy.
However, this cold layer, with respect to the \citet{lellouchetal2010} profile, is also robustly found in temperature retrievals with fixed CO (Fig. \ref{fig:appendix_TdifferentCO}), and is in fact key for accurately fitting the disk-average line profile.
A similar cold layer, in that case at about 53\,K and near 150\,mbar, was also derived in \citet{burgdorfetal2003} from disk-averaged observations by ISO in 1996 and 1997. 
In our case, however, the global temperature minimum in the atmosphere is located at significantly higher altitudes.

From these disk-integrated retrievals, we derived a global CO abundance profile.
We find distinctly different tropospheric and stratospheric abundances, with CO$_{tr}=0.27^{+0.20}_{-0.14}$ppm, CO$_{str}=0.69^{+0.27}_{-0.10}$ppm, and $P_0=135^{+184}_{-85}$mbar as the mean result of both retrievals with different temperature parameterizations.
Although the absolute values of CO are somewhat uncertain due to the T-CO degeneracy, the dual population of CO in the troposphere and stratosphere is confirmed.
This result is here obtained without any temperature profile assumptions, unlike in previous works
This validates the conclusions of \citet{lellouchetal2005}, who first reported the dual abundance of Neptune's CO and proposed that it has both internal and external sources, with the most likely external source being a cometary impact of 2\,km size about 200 years ago. From a similar study but using other data, \citet{hesmanetal2007} reported CO abundances of $2.2^{+0.6}_{-0.4}\times 10^{-6}$ in the upper stratosphere and $0.6\pm0.4\times 10^{-6}$ in the lower stratosphere and troposphere, and suggested an impact of a 12\,km-diameter comet about 100 years ago --a much more unlikely event that they estimate would only happen once every 55000 years.
Our retrieved CO values are more similar to those reported by \citet{lellouchetal2005}.
Still, using the above mean values for CO$_{tr}$, CO$_{str}$ and $P_0$, the stratospheric column density excess is $1.4\times10^{19}$\,cm$^{-2}$, 5.6 times larger than in \citet{lellouchetal2005}, and the timescale for diffusion down to 135\,mbar, where the estimated eddy $K$ coefficient is about 2000\,cm$^2$s$^{-1}$ \citep{romanietal1993}, is 280 years. 
Using the same approach as in \citet{lellouchetal2005}, this would imply, in this scenario, an impact of a $\sim$4\,km diameter comet 280\,yr ago. Once again, these numbers are indicative, given the CO-temperature degeneracy.

ALMA's high spatial resolution allowed us to measure spectra at 83 independent positions onto Neptune's 2.24" disk (Fig. \ref{fig:observations}).
By running spatially-resolved retrievals for each of the observed latitudes (-90$^\circ$ to +53$^\circ$) with a fixed, horizontally-uniform temperature profile, we found that CO latitudinal variations failed at reproducing the measured spectra (Table \ref{table:appendix_CO-only}, Fig. \ref{fig:appendix_onlyCO}), and implied a CO distribution inconsistent with current understanding of Neptune's equatorial stacked circulation.
This is in agreement with the conclusions of \citet{luszczcooketal2013} from mm CARMA observations of Neptune in the CO(2-1) line.
We therefore fixed a nominal CO profile of CO$_{tr}$=0.2\,ppm, CO$_{str}$=1.0\,ppm, and $P_0$=100\,mbar, as indicated for low latitudes.
We note that this profile is also consistent with those reported in \citet{lellouchetal2005}, and with the best-fitting CO profile in \citet{teanbyetal2019}.
With this nominal CO profile fixed throughout the planet, we carried out temperature retrievals at each latitude.

Spatially resolved temperature retrievals indicate a continuous variation of the thermal profile from pole to pole (Figs. \ref{fig:temperatures_multiplelatitudes}, \ref{fig:Tlatitude-cuts}, \ref{fig:Tmaps} and \ref{fig:appendix_retrievals8pointVS8order}).
In the lower stratosphere (10--40\,mbar), we find a temperature minimum at mid-latitudes, with the equator and south pole being warmer (e.g. Fig. \ref{fig:Tlatitude-cuts}).
We also find that this stratospheric temperature minimum --which can be observed in Figs. \ref{fig:temperatures_multiplelatitudes} and \ref{fig:appendix_retrievals8pointVS8order} as a deviation from the \citet{lellouchetal2010} reference profile-- varies in altitude, from about 20\,mbar at high latitudes, to slightly deeper levels (40\,mbar) at southern mid-latitudes (-60$^\circ$ to -20$^\circ$), until it disappears towards the equator.
In the upper troposphere (300--600\,mbar), however, we find the temperature minimum at southern polar latitudes instead of at mid-latitudes, with a cold spot of 45$\pm$3\,K at 400\,mbar at -80$^\circ$.

In general, our latitudinal thermal structure is in accordance with that derived by Voyager \citep{conrathetal1989, conrathetal1998}, and by subsequent ground-based mid-IR observations \citep[e.g.][]{fletcheretal2014, romanetal2022}.
In particular in the lower stratosphere (10--50\,mbar), our results are in agreement with the trends observed previously at the tropopause level (50-200 mbar).
At these levels we consistently find colder mid-latitudes, with warmer equator and south pole.
This suggests that if dynamics is indeed behind these trends (e.g. circulation cells), then these circulation patterns would need to extend to the lower stratosphere.
In our temperature maps (Fig. \ref{fig:Tmaps}), we observe a cold region at southern mid-latitudes between about 40 and 200\,mbar. This structure was also identified by \citet[][see Fig. 18 therein]{fletcheretal2014} and \citet[][see Fig. 20 therein]{romanetal2022}, although we find it to be more extended fand isothermal.
\citet{romanetal2022} found this cold region in all of their 2003-2020 observations, although they reported slight variations between the different datasets, suggesting subseasonal variability to happen also at the tropopause level.

At deeper levels, however, we find some differences with respect to previous thermal fields in the literature.
Most notably, we report here a southern polar cold spot of 45$\pm$3\,K at about 400\,mbar. This cold spot fades away towards mid-latitudes and the equator.
This feature is well outside of our retrieval errors (3\,K at -80$^\circ$) and also cannot be explained by potential systematic uncertainties.
As mentioned in Sect. \ref{sec:observations}, the absolute flux scale is estimated to be accurate to within 5\%. This typically corresponds to a 2.5\,K absolute error on a 50\,K temperature, but the latter would be systematic and not cause a specific structure at some latitude.
\citet{fletcheretal2014} did report a hint for a lower 200\,mbar temperature near -85$^\circ$ latitudes, although they suggested that it could be due to spurious projection effects.
\citet{romanetal2022} also found that, at 50--300\,mbar levels, the south pole was 6$\pm$3\,K colder in 2018--2020 than in 2003--2006, although their polar temperatures at 100\,mbar never reached below 52\,K (their Fig. 22).
Similarly, we find our 100\,mbar south pole temperatures to be on the order of 52\,K, but they decrease significantly at 400\,mbar (Fig. \ref{fig:Tlatitude-cuts}).
We argue that the sensitivity of our data to deeper layers and to cold temperatures is the main reason why we detected this previously unreported cold spot.
We note that we carried out additional retrievals --e.g., polynomial retrievals with different maximum pressures probed (1\,bar and 10\,bar), as well as independent-point retrievals with more sampling points. All of them showed consistent results, including this cold polar anomaly at the same pressure levels.

Despite these differences with previous thermal fields, the computed thermal winds, especially in the best characterized southern hemisphere, show similar structure to those from \citet[][their Fig. 18]{fletcheretal2014}. They both show essentially pressure-independent zonal winds, i.e. small wind shears. This behaviour is not fully consistent with the fact that directly measured Doppler winds at 0.4-2 mbar \citep{carriongonzalezetal2023} are weaker than Voyager-measured cloud top winds. We note however, that increasing temperature error bars at P\,<\,10\,mbar make the derivation of thermal winds in the mbar region accordingly uncertain.

Characterizing the latitudinal thermal profiles of Neptune in an extended pressure range is particularly useful to understand the global circulation of the planet and validate general circulation models \citep[e.g.][]{milcarecketal2024}.
For instance, \citet{depateretal2014} derived a global atmospheric circulation pattern with two cells per hemisphere, in which air rises from mid-latitudes and is subducted both at the poles and at the equator.
They observed the previously reported south-polar brightness enhancement \citep{ortonetal2007, ortonetal2012} in mid-IR images (8--22\,$\mu$m, probing 0.1--200\,mbar), and also VLA radio measurements (0.7--6.2\,cm, probing 5--50\,bar).
They interpreted this as evidence of stratospheric air being subducted down to the deep troposphere.
\citet{iinoetal2018} did not detect the southern polar hot spot from spatially resolved measurements of Neptune's continuum at 646\,GHz with ALMA, with an uncertainty of about 2\,K.
They found that these observations probed at pressures of about 0.6--1.0\,bar, and suggested that wind shear at south polar regions could be the cause of the polar hot spot not being continuous in altitude. 
A more recent view, indeed, is that the circulation cells extend only down to about 1\,bar, with inverted cells at deeper levels \citep[see][their Fig. 5]{fletcheretal2020}.
Our results for 0.2--1.0\,bar indicate that this discontinuity region reaches even higher atmospheric layers, and that it contains a cold spot at 0.4\,bar. 
This structure resembles that observed in Jupiter by Juno's microwave radiometer \citep{boltonetal2021}, where the vortices are found to have a complex structure alternating colder and warmer layers.

Neptune's south polar region has proved to be highly variable, with brightness peaks \citep{ortonetal2007, ortonetal2012, depateretal2014}, subseasonal temperature variations \citep{fletcheretal2014, romanetal2022}, and meteorological activity such as the South Polar Feature \citep{sromovskyetal1993, ragesetal2002} and other deeper clouds potentially linked to convective activity of those latitudes \citep[e.g.][]{irwinetal2016}.
The single epoch studied in this work is not enough to determine whether our reported cold spot is linked to this variability, or if it is a stable feature of the south polar region at 400\,mbar.
Intensive monitoring of the southern polar region with spatially-resolved measurements sensitive to upper tropospheric levels is needed to complete the picture of this active region of Neptune before it becomes unobservable from the Earth as Neptune orbits towards its northern summer solstice in 2087, with its northern spring equinox in 2046.

\begin{acknowledgements}

This paper makes use of the following ALMA data:
ADS/JAO.ALMA\#2015.1.01471.S. (PI: R. Moreno). ALMA is a partnership of ESO (representing its member states), NSF (USA) and NINS (Japan), together with NRC (Canada), NSC and ASIAA (Taiwan), and KASI (Republic of Korea), in cooperation with the Republic of Chile.
The Joint ALMA Observatory is operated by ESO, AUI/NRAO and NAOJ. 
The authors acknowledge the support of the French Agence Nationale de la Recherche (ANR), under grant ANR-20-CE49-0009 (project SOUND).
The authors thank A. Spiga for the helpful discussions within project SOUND.
T. Cavali\'e acknowledges funding from CNES.
\end{acknowledgements}

%
%

\bibliographystyle{aa}
\bibliography{references}
\clearpage

\begin{appendix} 

\onecolumn

\begin{figure*}
\section{Disk-averaged retrievals at different fixed CO profiles} \label{sec:appendix_diskav_differentCO}
    \centering
	\includegraphics[width=12.5cm]{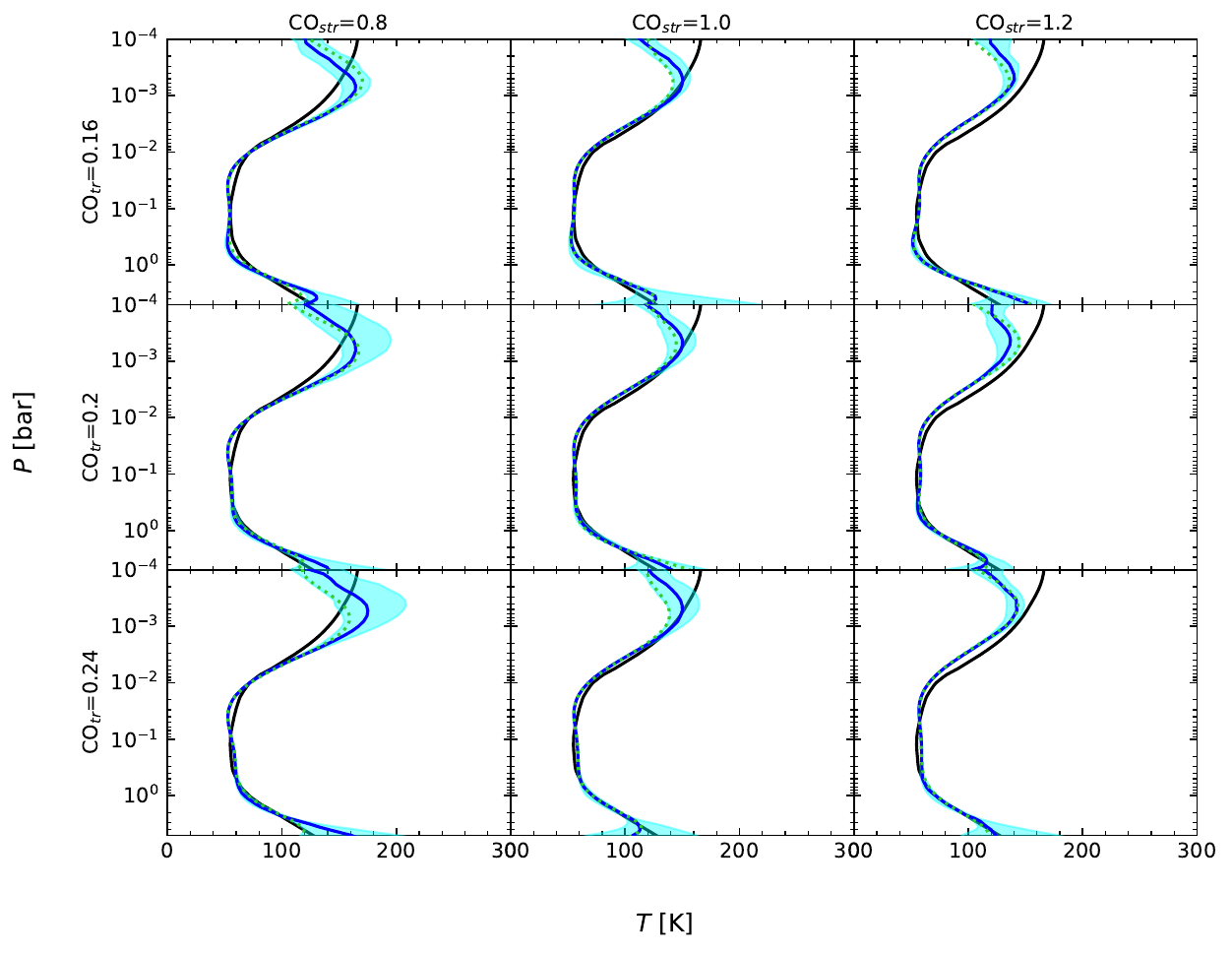}
    \caption{Retrieved temperature profiles for different combinations of fixed CO profiles with P$_0$ = 100\,mbar. As in Figs. \ref{fig:results_TP+CO_poly8}, \ref{fig:results_TP+CO_8point} and \ref{fig:temperatures_multiplelatitudes}, blue lines indicate the median of the ensemble of good fits, green lines (left) or dots (right) represent the best-fit profile, and black lines show the \citet{lellouchetal2010} profile for reference.}
\label{fig:appendix_TdifferentCO}
\end{figure*}

\section{CO-only retrievals} \label{sec:appendix_CO-only} 

We report in Table \ref{table:appendix_CO-only} CO-only retrievals for a fixed temperature profile. This was performed by using the \citet{lellouchetal2010} temperature profile. Arguably, this profile may not be representative of Neptune's mean conditions in 2016. However, we repeated the same exercise by using instead the best-fit temperature profile from the coupled CO/temperature inversion of the disk-averaged spectrum (independent-point parameterization, green dots in Fig. \ref{fig:results_TP+CO_8point}). For obvious circularity reasons, the retrieved disk-averaged CO profile was identical to that of Fig. \ref{fig:results_TP+CO_8point} (green line), with a good fit, but the inferred CO latitudinal distribution continued to show large variations in the CO parameters, particularly P$_0$, which fits comparable to those of this Table \ref{table:appendix_CO-only}.

\begin{table*}[ht!]
\begin{center}
\caption{Results of CO-only retrievals at different latitudes, keeping in all cases the temperature profile fixed to \citet{lellouchetal2010}. }
\label{table:appendix_CO-only} 
\begin{tabular}{c c c c c c}
\hline 
\hline
  & \multicolumn{4}{c}{CO-only retrievals} & T-only retrieval \\
Latitude &  CO$_{tr}$[ppm]    &     CO$_{str}$[ppm]     & $P_{0}$[mbar]   &  $\chi^2_{min}/N$  & $\chi^2_{min}/N$ \\
\hline
 $-90^\circ$ & $0.43^{+0.03}_{-0.05}$  & $0.85^{+0.09}_{-0.08}$  & $88^{+41}_{-41}$  & 1.22  & 1.48 \\
 $-80^\circ$ & $0.38^{+0.05}_{-0.07}$  & $0.80^{+0.09}_{-0.08}$  & $109^{+56}_{-58}$  & 9.92 & 3.20 \\
 $-64^\circ$ & $0.19^{+0.03}_{-0.05}$  & $0.82^{+0.07}_{-0.06}$  & $150^{+28}_{-25}$  & 6.32 & 2.95 \\
 $-52^\circ$ & $0.15^{+0.04}_{-0.05}$  & $0.79^{+0.07}_{-0.06}$  & $167^{+31}_{-29}$  & 7.32 & 2.49\\
 $-43^\circ$ & $0.17^{+0.02}_{-0.03}$  & $0.88^{+0.06}_{-0.07}$  & $144^{+21}_{-21}$  & 7.44 & 2.44 \\
 $-32^\circ$ & $0.12^{+0.02}_{-0.02}$  & $0.96^{+0.07}_{-0.05}$  & $132^{+17}_{-11}$   & 4.73 & 2.46 \\
 $-20^\circ$ & $0.19^{+0.02}_{-0.05}$  & $1.07^{+0.11}_{-0.12}$  & $107^{+18}_{-8}$  & 3.24 & 2.45 \\
 $-10^\circ$ & $0.18^{+0.02}_{-0.02}$  & $1.07^{+0.08}_{-0.08}$  & $96^{+12}_{-11}$  & 2.66 & 2.33 \\
 $0^\circ$ & $0.17^{+0.02}_{-0.02}$  & $1.04^{+0.09}_{-0.08}$  & $85^{+12}_{-14}$  & 3.01 & 2.64 \\
 $+12^\circ$ & $0.15^{+0.02}_{-0.02}$  & $1.00^{+0.08}_{-0.07}$  & $96^{+12}_{-11}$  & 2.70 & 1.78 \\
 $+26^\circ$ & $0.09^{+0.02}_{-0.02}$  & $0.82^{+0.06}_{-0.05}$  & $132^{+17}_{-12}$  & 5.86 & 1.74 \\
 $+37^\circ$ & $0.09^{+0.06}_{-0.04}$  & $0.66^{+0.05}_{-0.04}$  & $192^{+37}_{-28}$  & 9.97 & 1.51 \\
 $+53^\circ$ & $0.09^{+0.06}_{-0.04}$  & $0.61^{+0.05}_{-0.05}$  & $225^{+41}_{-35}$  & 13.5 & 2.03 \\
\hline
\end{tabular}
\end{center}
\tablefoot{For comparison, the last column shows the best-fit reduced $\chi^2$ for temperature-only retrievals (polynomial parameterization of T) with a fixed CO profile ( CO$_{tr}$=0.2\,ppm, CO$_{str}$=1.0\,ppm, and $P_0$=100\,mbar).}
\end{table*}

\begin{figure*}[ht!]
  \centering
  \includegraphics[width=9cm]{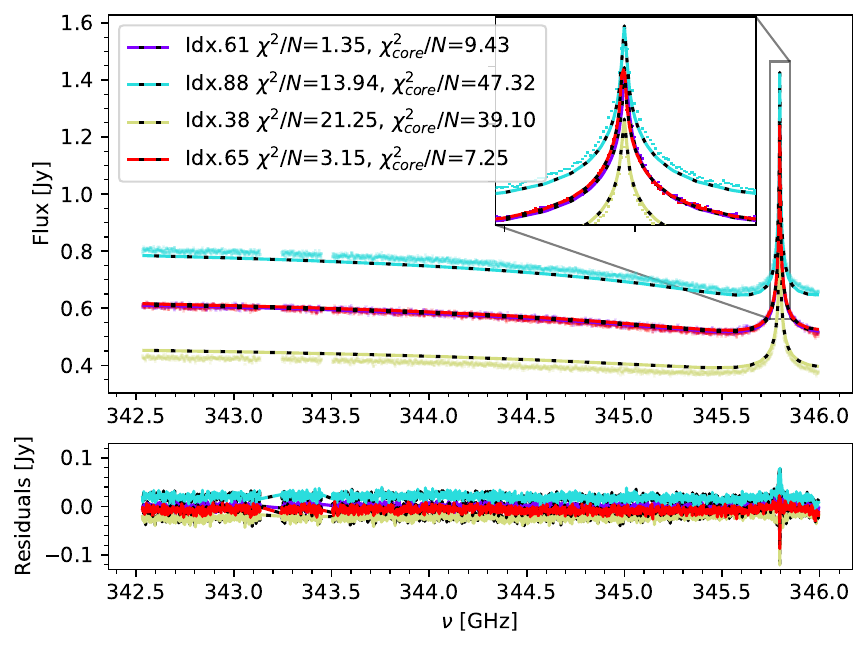}
  \hfill
  \includegraphics[width=9cm]{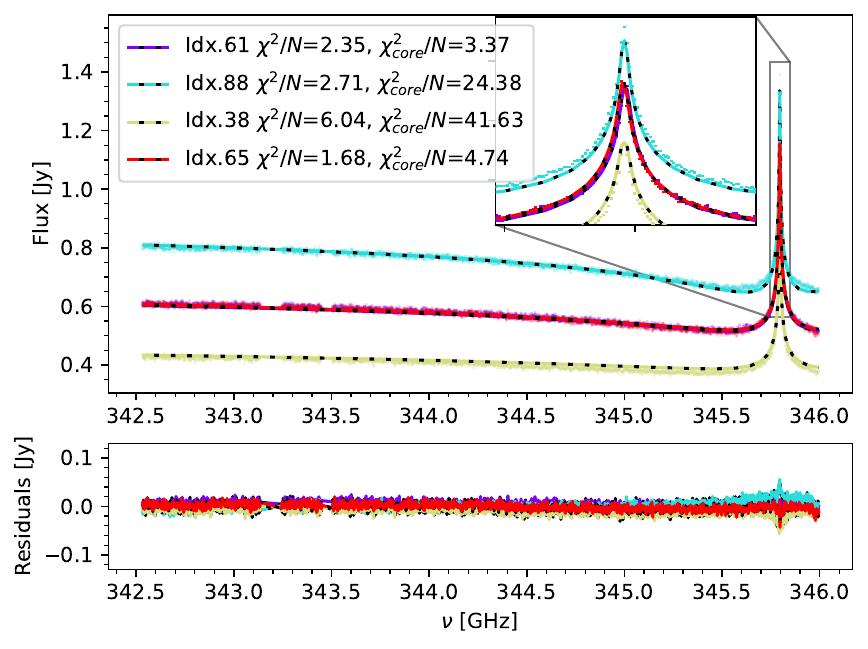}
    \caption{Best-fitting spectra for the CO-only retrievals with a fixed temperature profile (left), and for the temperature-only retrievals with a fixed CO profile (right), as in Table \ref{table:appendix_CO-only}, at an example latitude of -80$^\circ$. For each of the four beam positions observed at that latitude (Fig. \ref{fig:observations}) --each of them labeled with a different index--, a synthetic spectrum was generated from the best-fit atmospheric configuration, and compared to the measured spectrum. The corresponding residuals are also shown.}
\label{fig:appendix_onlyCO}
\end{figure*}

\begin{figure*}
    \section{Latitude-dependent temperature retrievals} \label{sec:appendix_retrievals8pointVS8order}
    \includegraphics[width=15cm]{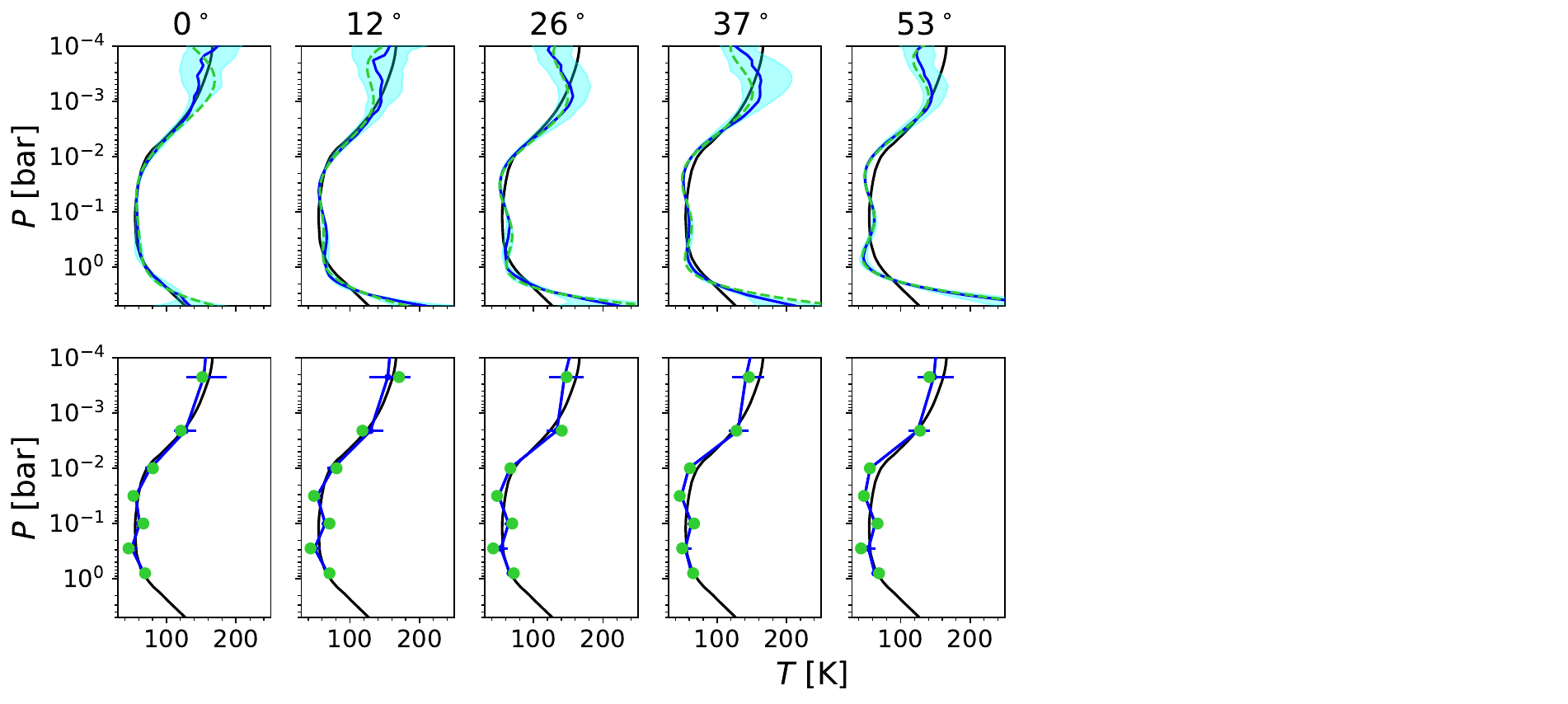}
    \includegraphics[width=15cm]{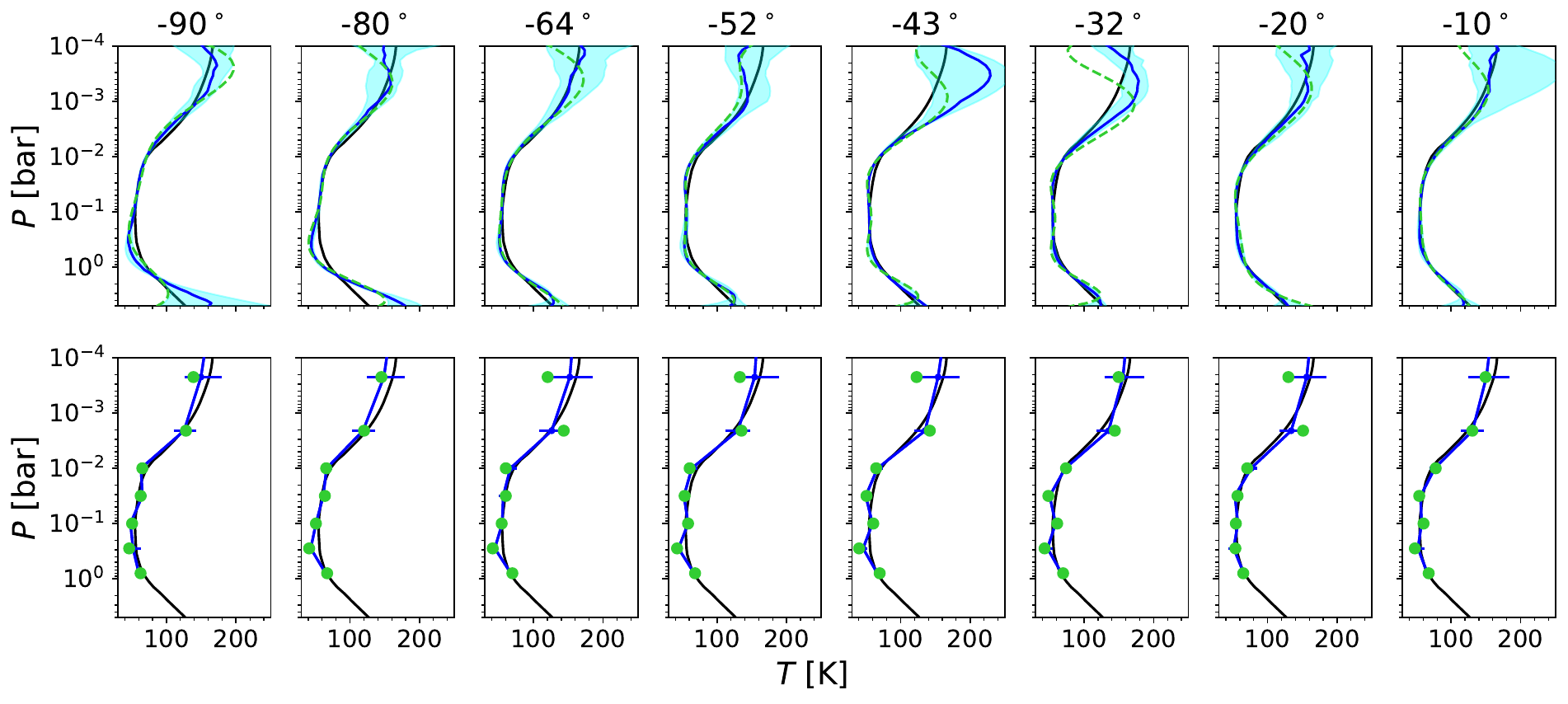}
    \caption{As Fig. \ref{fig:temperatures_multiplelatitudes}, but here showing the whole set of latitudes for which temperature-only retrievals carried out. Two pairs of rows are shown, for Northern latitudes (up) and Southern latitudes (down). In each case, both the 8th order polynomial parameterization (up) and the independent-point parameterization (down) are included.}
\label{fig:appendix_retrievals8pointVS8order}
\end{figure*}

\end{appendix}

\end{document}